\documentclass[11pt]{article}
\usepackage[dvips]{graphicx}
\usepackage{float}
\usepackage{epsfig}
\usepackage{latexsym,amsmath,amsfonts,amssymb}
\usepackage{rotating}
\usepackage[american]{babel}
\usepackage[dvips]{graphicx}
\usepackage{bbm}
\usepackage{color}
\usepackage{slashed}
\usepackage[unicode]{hyperref}
\usepackage{lscape}
\usepackage{bigints}
\usepackage{enumerate}
\usepackage[shortlabels]{enumitem}
\usepackage{subcaption}
\usepackage{tikz}
\usetikzlibrary{decorations.pathreplacing}
\usetikzlibrary{shapes}
\usepackage{graphicx}
\usepackage{epsfig}
\usepackage{rotating}
\usepackage{amssymb}
\usepackage{dsfont}
\usepackage{psfrag}
\usepackage{amsmath,euscript,array,mathrsfs,amsfonts}
\usepackage{slashed}
\usepackage{array}
\usepackage{youngtab}
\usepackage{color}
\usepackage{bbold}
\usepackage{hyperref}
\hypersetup{
colorlinks = true,
linkcolor = red,
linktocpage = true,
citecolor = blue
}

\usepackage{tikz}
\usetikzlibrary{calc}
 \usetikzlibrary{decorations.text}
 \usetikzlibrary{shapes}

 \usetikzlibrary{decorations.pathmorphing}
\usetikzlibrary{decorations.pathreplacing}
\usetikzlibrary{arrows.meta}
\tikzset{
  >={To[length=5pt]}
  }
\usetikzlibrary{shapes, shapes.geometric, shapes.symbols, shapes.arrows, shapes.multipart, shapes.callouts, shapes.misc}
\tikzset{snake it/.style={decorate, decoration=snake}}
\tikzset{7brane/.style={circle, draw=black, fill=black,ultra thick,inner sep=1.5 pt, minimum size=1 pt,}, c/.default={4pt}}
\tikzset{cross/.style={cross out, draw=black,thick, minimum size=2*(#1-\pgflinewidth), inner sep=0pt, outer sep=0pt}, cross/.default={5pt}}
\tikzset{big7brane/.style={circle, draw=black, fill=black,ultra thick,inner sep=2.5 pt, minimum size=1 pt,}, c/.default={4pt}}
\tikzset{u/.style={circle, draw=black, fill=white,inner sep=2 pt, minimum size=2 pt,},f/.style={square, draw=black, fill=white,ultra thick,inner sep=4 pt, minimum size=2 pt,}}
\tikzset{so/.style={circle, draw=black, fill=red,inner sep=2 pt, minimum size=2 pt,},f/.style={square, draw=black, fill=white,ultra thick,inner sep=4 pt, minimum size=2 pt,}}
\tikzset{sp/.style={circle, draw=black, fill=blue,inner sep=2 pt, minimum size=2 pt,},f/.style={square, draw=black, fill=white,ultra thick,inner sep=4 pt, minimum size=2 pt,}}
\tikzset{uf/.style={rectangle, draw=black, fill=white,inner sep=3 pt, minimum size=4 pt,}}
\tikzset{spf/.style={rectangle, draw=black, fill=blue, thick,inner sep=3 pt, minimum size=4 pt, circle, draw=black, fill=blue,thick,inner sep=2 pt, minimum size=2 pt,},f/.style={square, draw=black, fill=white,ultra thick,inner sep=4 pt, minimum size=2 pt,}}
\tikzset{sof/.style={rectangle, draw=black, fill=red, thick,inner sep=3 pt, minimum size=4 pt,}}
\usetikzlibrary{positioning}
\usetikzlibrary{arrows}
\usetikzlibrary{decorations.pathreplacing}
\usetikzlibrary{shapes}

\makeatletter\def\l@subsubsection#1#2{}%
\makeatother

\renewcommand\theequation{\arabic{section}.\arabic{equation}} 

\def\cA{{\cal A}}

\def\cF{{\cal F}}

\def\CC{\ensuremath{\mathds C}}
\def\RR{\ensuremath{\mathds R}}
\def\ZZ{\ensuremath{\mathds Z}}

\DeclareMathOperator{\vol}{vol}

\DeclareMathOperator{\sech}{sech}
\DeclareMathOperator{\tr}{tr}

\DeclareMathOperator{\csch}{csch}

\newcommand{\be}{\begin{equation}}
\newcommand{\ee}{\end{equation}}
\newcommand{\ba}{\begin{array}}
\newcommand{\ea}{\end{array}}

\def\im{Invent. Math.}

\def\hat{\widehat}
\def\a{\alpha}
\def\b{\beta}
\def\c{\gamma}
\def\d{\delta}
\def\f{\phi}               
\def\vf{\varphi}  
\def\tvf{\tilde{\varphi}}
\def\vp{\varphi}
\def\g{\gamma}
\def\h{\eta}
\def\j{\psi}
\def\k{\kappa}                    
\def\l{\lambda}
\def\m{\mu}
\def\n{\nu}
\def\o{\omega}  \def\w{\omega}
\def\p{\pi}
\def\q{\theta}  \def\th{\theta}                  
\def\r{\rho}                                     
\def\s{\sigma}                                   
\def\t{\tau}
\def\u{\upsilon}
\def\x{\xi}
\def\z{\zeta}
\def\pt{\tilde{\varphi}}
\def\tt{\tilde{\theta}}
\def\lab{\label}
\def\6{\partial}
\def\wg{\wedge}
\def\bpsi{\bar{\psi}}
\def\bt{\bar{\theta}}
\def\bvf{\bar{\varphi}}
\def\W{\Omega}

\newcommand{\nn}{\nonumber \\}
\newcommand{\td}{\mathrm{d}}

\DeclareMathOperator{\str}{str}

\newcommand{\beq}{\begin{equation}}
\newcommand{\eeq}{\end{equation}}
\newcommand{\bea}{\begin{eqnarray}}
\newcommand{\eea}{\end{eqnarray}}

\newcommand{\beqs}{\begin{eqnarray}}
\newcommand{\eeqs}{\end{eqnarray}}
\newcommand{\bal}{\begin{aligned}}
\newcommand{\eal}{\end{aligned}}
\makeatletter
\newcommand\setItemnumber[1]{\setcounter{enum\romannumeral\@enumdepth}{\numexpr#1-1\relax}}
\makeatother
\begin{document}
\baselineskip=15.5pt
\pagestyle{plain}
\setcounter{page}{1}

\def\del{{\partial}}
\def\vev#1{\left\langle #1 \right\rangle}
\def\cn{{\cal N}}
\def\co{{\cal O}}


\def\IC{{\mathbb C}}
\def\IR{{\mathbb R}}
\def\IZ{{\mathbb Z}}
\def\RP{{\bf RP}}
\def\CP{{\bf CP}}
\def\Poincaré{{Poincar\'e }}
\def\tr{{\rm tr}}
\def\tp{{\tilde \Phi}}

\def\TL{\hfil$\displaystyle{##}$}
\def\TR{$\displaystyle{{}##}$\hfil}
\def\TC{\hfil$\displaystyle{##}$\hfil}
\def\TT{\hbox{##}}
\def\HLINE{\noalign{\vskip1\jot}\hline\noalign{\vskip1\jot}}
\def\seqalign#1#2{\vcenter{\openup1\jot
   \halign{\strut #1\cr #2 \cr}}}
\def\lbldef#1#2{\expandafter\gdef\csname #1\endcsname {#2}}
\def\eqn#1#2{\lbldef{#1}{(\ref{#1})}%
\begin{equation} #2 \label{#1} \end{equation}}
\def\eqalign#1{\vcenter{\openup1\jot
     \halign{\strut\span\TL & \span\TR\cr #1 \cr
    }}}

\def\eno#1{(\ref{#1})}
\def\href#1#2{#2}
\def\half{\frac{1}{2}}



\def\ads{{\it AdS}}
\def\adsp{{\it AdS}$_{p+2}$}
\def\cft{{\it CFT}}

\newcommand{\ber}{\begin{eqnarray}}
\newcommand{\eer}{\end{eqnarray}}

\newcommand{\beqar}{\begin{eqnarray}}
\newcommand{\cO}{{\cal O}}
\newcommand{\cT}{{\cal T}}
\newcommand{\cR}{{\cal R}}
\newcommand{\eeqar}{\end{eqnarray}}
\newcommand{\tht}{\thteta}
\newcommand{\lm}{\lambda}\newcommand{\Lm}{\Lambda}


\newcommand{\nonu}{\nonumber}
\newcommand{\oh}{\displaystyle{\frac{1}{2}}}
\newcommand{\dsl}
   {\kern.06em\hbox{\raise.15ex\hbox{$/$}\kern-.56em\hbox{$\partial$}}}
\newcommand{\as}{\not\!\! A}
\newcommand{\ps}{\not\! p}
\newcommand{\ks}{\not\! k}
\newcommand{\D}{{\cal{D}}}
\newcommand{\dv}{d^2x}
\newcommand{\Z}{{\cal Z}}
\newcommand{\N}{{\cal N}}
\newcommand{\Dsl}{\not\!\! D}
\newcommand{\Bsl}{\not\!\! B}
\newcommand{\Psl}{\not\!\! P}

\newcommand{\eeqarr}{\end{eqnarray}}


\def\del{{\delta^{\hbox{\sevenrm B}}}} \def\ex{{\hbox{\rm e}}}
\def\azb{A_{\bar z}} \def\az{A_z} \def\bzb{B_{\bar z}} \def\bz{B_z}
\def\czb{C_{\bar z}} \def\cz{C_z} \def\dzb{D_{\bar z}} \def\dz{D_z}
\def\im{{\hbox{\rm Im}}} \def\mod{{\hbox{\rm mod}}} \def\tr{{\hbox{\rm Tr}}}
\def\ch{{\hbox{\rm ch}}} \def\imp{{\hbox{\sevenrm Im}}}
\def\trp{{\hbox{\sevenrm Tr}}} \def\vol{{\hbox{\rm Vol}}}
\def\rl{\Lambda_{\hbox{\sevenrm R}}} \def\wl{\Lambda_{\hbox{\sevenrm W}}}
\def\fc{{\cal F}_{k+\cox}} \def\vev{vacuum expectation value}
\def\nodiv{\mid{\hbox{\hskip-7.8pt/}}}
\def\ie{{\em i.e.}}
\def\ie{\hbox{\it i.e.}}

\def\CC{{\mathchoice
{\rm C\mkern-8mu\vrule height1.45ex depth-.05ex
width.05em\mkern9mu\kern-.05em}
{\rm C\mkern-8mu\vrule height1.45ex depth-.05ex
width.05em\mkern9mu\kern-.05em}
{\rm C\mkern-8mu\vrule height1ex depth-.07ex
width.035em\mkern9mu\kern-.035em}
{\rm C\mkern-8mu\vrule height.65ex depth-.1ex
width.025em\mkern8mu\kern-.025em}}}

\def\RR{{\rm I\kern-1.6pt {\rm R}}}
\def\NN{{\rm I\!N}}
\def\ZZ{{\rm Z}\kern-3.8pt {\rm Z} \kern2pt}
\def\IB{\relax{\rm I\kern-.18em B}}
\def\ID{\relax{\rm I\kern-.18em D}}
\def\II{\relax{\rm I\kern-.18em I}}
\def\IP{\relax{\rm I\kern-.18em P}}
\newcommand{\CS}{{\scriptstyle {\rm CS}}}
\newcommand{\CSs}{{\scriptscriptstyle {\rm CS}}}
\newcommand{\rc}{\nonumber\\}
\newcommand{\bear}{\begin{eqnarray}}
\newcommand{\eear}{\end{eqnarray}}

\newcommand{\LL}{{\cal L}}

\def\mani{{\cal M}}
\def\calo{{\cal O}}
\def\calb{{\cal B}}
\def\calw{{\cal W}}
\def\calz{{\cal Z}}
\def\cald{{\cal D}}
\def\calc{{\cal C}}
\newcommand{\gt}{\tilde{g}}

\def\to{\rightarrow}
\def\ele{{\hbox{\sevenrm L}}}
\def\ere{{\hbox{\sevenrm R}}}
\def\zb{{\bar z}}
\def\wb{{\bar w}}
\def\nodiv{\mid{\hbox{\hskip-7.8pt/}}}
\def\menos{\hbox{\hskip-2.9pt}}
\def\dr{\dot R_}
\def\drr{\dot r_}
\def\ds{\dot s_}
\def\da{\dot A_}
\def\dga{\dot \gamma_}
\def\ga{\gamma_}
\def\dal{\dot\alpha_}
\def\al{\alpha_}
\def\cl{{closed}}
\def\cls{{closing}}
\def\vev{vacuum expectation value}
\def\tr{{\rm Tr}}
\def\to{\rightarrow}
\def\too{\longrightarrow}

\newcommand{\dd}{\mathrm{d}}

\def\a{\alpha}
\def\b{\beta}
\def\c{\gamma}
\def\d{\delta}
\def\e{\epsilon}           
\def\F{\Phi}
\def\f{\phi}               
\def\vf{\varphi}  \def\tvf{\tilde{\varphi}}
\def\vp{\varphi}
\def\g{\gamma}
\def\h{\eta}
\def\j{\psi}
\def\k{\kappa}                    
\def\l{\lambda}
\def\m{\mu}
\def\n{\nu}
\def\o{\omega}  \def\w{\omega}
\def\q{\theta}  \def\th{\theta}                  
\def\r{\rho}                                     
\def\s{\sigma}                                   
\def\t{\tau}
\def\u{\upsilon}
\def\x{\xi}
\def\X{\Xi}
\def\z{\zeta}
\def\pt{\tilde{\varphi}}
\def\tt{\tilde{\theta}}
\def\lab{\label}
\def\6{\partial}
\def\wg{\wedge}
\def\atanh{{\rm arctanh}}
\def\bpsi{\bar{\psi}}
\def\bt{\bar{\theta}}
\def\bvf{\bar{\varphi}}

\def\ft#1#2{{\textstyle{{\scriptstyle #1}\over {\scriptstyle #2}}}}
\def\fft#1#2{{#1 \over #2}}
\def\del{\partial}
\def\sst#1{{\scriptscriptstyle #1}}

\def\dalemb#1#2{{\vbox{\hrule height .#2pt
        \hbox{\vrule width.#2pt height#1pt \kern#1pt
                \vrule width.#2pt}
        \hrule height.#2pt}}}
\def\square{\mathord{\dalemb{6.8}{7}\hbox{\hskip1pt}}}
\def\hF{\hat F}
\def\tA{\widetilde A}
\def\tcA{{\widetilde{\cal A}}}
\def\tcF{{\widetilde{\cal F}}}
\def\hA{\hat{\cal A}}
\def\cF{{\cal F}}
\def\cA{{\cal A}}
\def\wdg{{\sst \wedge}}

\def\0{{\sst{(0)}}}
\def\1{{\sst{(1)}}}
\def\2{{\sst{(2)}}}
\def\3{{\sst{(3)}}}
\def\4{{\sst{(4)}}}
\def\5{{\sst{(5)}}}
\def\6{{\sst{(6)}}}
\def\7{{\sst{(7)}}}
\def\8{{\sst{(8)}}}
\def\n{{\sst{(n)}}}
\def\tV{\widetilde V}
\def\tW{\widetilde W}
\def\tH{\widetilde H}
\def\tE{\widetilde E}
\def\tF{\widetilde F}
\def\tA{\widetilde A}
\def\tP{{\widetilde P}}
\def\tD{\widetilde D}
\def\bA{\bar{\cal A}}
\def\bF{\bar{\cal F}}
\def\tG{\widetilde G}
\def\tT{\widetilde T}
\def\Z{\rlap{\sf Z}\mkern3mu{\sf Z}}
\def\R{\rlap{\rm I}\mkern3mu{\rm R}}
\def\G{{\cal G}}
\def\gg{\bf g}
\def\CS{{\cal S}}
\def\S{{\cal S}}
\def\P{{\cal P}}
\def\ep{\epsilon}
\def\td{\tilde}
\def\wtd{\widetilde}
\def\half{{\textstyle{1\over2}}}
\def\Qw{{Q_{\rm wave}}}
\def\Qnut{{Q_{\sst{\rm NUT}}}}
\def\mun{{\mu_{\sst{\rm NUT}}}}
\def\muw{{\mu_{\rm wave}}}
\let\a=\alpha \let\b=\beta \let\g=\gamma \let\d=\delta \let\e=\epsilon
\let\z=\zeta \let\h=\eta \let\q=\theta \let\i=\iota \let\k=\kappa
\let\l=\lambda \let\m=\mu \let\n=\nu \let\x=\xi \let\p=\pi \let\r=\rho
\let\s=\sigma \let\t=\tau \let\u=\upsilon \let\f=\phi \let\c=\chi \let\y=\psi
\let\w=\omega  \let\D=\Delta \let\Q=\Theta \let\L=\Lambda
\let\X=\Xi  \let\U=\Upsilon \let\F=\Phi \let\Y=\Psi
\let\C=\Chi \let\W=\Omega     
\let\la=\label \let\ci=\cite \let\re=\ref
\let\se=\section \let\sse=\subsection \let\ssse=\subsubsection 
\def\bd{\begin{document}} \def\ed{\end{document}}
\def\ds{\documentstyle} \let\fr=\frac \let\bl=\bigl \let\br=\bigr
\let\Br=\Bigr \let\Bl=\Bigl 
\let\bm=\bibitem
\let\na=\nabla
\let\pa=\partial \let\ov=\overline 
\def\ba{\begin{eqnarray}}
\def\ea{\end{eqnarray}}
\def\ft#1#2{{\textstyle{{\scriptstyle #1}\over {\scriptstyle #2}}}}
\def\fft#1#2{{#1 \over #2}}
\def\del{\partial}
\def\sst#1{{\scriptscriptstyle #1}}
\def\oneone{\rlap 1\mkern4mu{\rm l}}
\def\ie{{\it i.e.\ }}
\def\via{{\it via}}
\def\semi{{\ltimes}}
\def\str{{\rm str}}
\def\jm{{\rm j}}
\def\im{{\rm i}}
\def\mapright#1{\smash{\mathop{-\!\!\!-\!\!\!-\!\!\!-\!\!\!-\!\!\!
             \longrightarrow}\limits^{#1}}}
\def\maprightt#1#2{\smash{\mathop{-\!\!\!-\!\!\!-\!\!\!-\!\!\!-\!\!\!
             \longrightarrow}\limits^{#1}_{#2}}}

\newcommand{\ho}[1]{$\, ^{#1}$}
\newcommand{\hoch}[1]{$\, ^{#1}$}
\newcommand{\ra}{\rightarrow}
\newcommand{\lra}{\longrightarrow}
\newcommand{\Lra}{\Leftrightarrow}
\newcommand{\bp}{\tilde \beta^\prime}
\newcommand{\Tr}{{\rm Tr} } 
\def\rme{{\rm e}}


\newfont{\namefont}{cmr10}
\newfont{\addfont}{cmti7 scaled 1440}
\newfont{\boldmathfont}{cmbx10}
\newfont{\headfontb}{cmbx10 scaled 1728}





\newcommand{\hyph}[1]{$#1$\nobreakdash-\hspace{0pt}}
\providecommand{\abs}[1]{\lvert#1\rvert}
\newcommand{\Nugual}[1]{$\mathcal{N}= #1 $}
\newcommand{\sub}[2]{#1_\text{#2}}
\newcommand{\partfrac}[2]{\frac{\partial #1}{\partial #2}}
\newcommand{\bsp}[1]{\begin{equation} \begin{split} #1 \end{split} \end{equation}}
\newcommand{\calF}{\mathcal{F}}
\newcommand{\calO}{\mathcal{O}}
\newcommand{\calM}{\mathcal{M}}
\newcommand{\calV}{\mathcal{V}}
\newcommand{\bbZ}{\mathbb{Z}}
\newcommand{\bbC}{\mathbb{C}}
\newcommand{\cK}{{\cal K}}

\newcommand{\Thq}{\Theta\left(\r-\r_q\right)}
\newcommand{\Dq}{\d\left(\r-\r_q\right)}
\newcommand{\kten}{\kappa^2_{\left(10\right)}}
\newcommand{\pbi}[1]{\imath^*\left(#1\right)}
\newcommand{\tth}{\tilde{\th}}
\newcommand{\tf}{\tilde{\f}}
\newcommand{\tj}{\tilde{\j}}
\newcommand{\tw}{\tilde{\omega}}
\newcommand{\tz}{\tilde{z}}
\newcommand{\prj}[2]{(\partial_r{#1})(\partial_{\j}{#2})-(\partial_r{#2})(\partial_{\j}{#1})}
\def\atanh{{\rm arctanh}}
\def\sech{{\rm sech}}
\def\csch{{\rm csch}}
\allowdisplaybreaks[1]

\def\red{\textcolor[rgb]{0.98,0.00,0.00}}

\newcommand{\Dan}[1] {{\textcolor{blue}{#1}}}

\numberwithin{equation}{section}



%

%
\setcounter{footnote}{0}
\renewcommand{\theequation}{{\rm\thesection.\arabic{equation}}}

\begin{titlepage}

\begin{center}

\vskip .5in 
\noindent

{\Large \bf{ Conformal to confining SQFTs from holography} }
\bigskip\medskip

Dimitrios Chatzis\footnote{d.chatzis.2322097@swansea.ac.uk}, Ali Fatemiabhari\footnote{a.fatemiabhari.2127756@swansea.ac.uk}, Carlos Nunez\footnote{c.nunez@swansea.ac.uk} and Peter Weck\footnote{p.j.weck@swansea.ac.uk} \\

\bigskip\medskip
{\small 
Department of Physics, Swansea University, Swansea SA2 8PP, United Kingdom}

\vskip .5cm 
\vskip .9cm 
     	{\bf Abstract }\vskip .1in
\end{center}

\noindent
In this paper we present three new families of smooth Type II string theory backgrounds. These are dual to supersymmetry-preserving deformations of 4d SCFTs. The deformations include a VEV for a global current and a `twisted compactification' on a circle. We study various holographic aspects of the dual QFTs, focusing on Wilson loops and Entanglement Entropy. Additionally, we present a monotonic quantity calculating the density of degrees of freedom in terms of the energy, which interpolates between the IR 3d gapped theory and the 4d SCFT result. Other probes related to global aspects of the QFTs are briefly discussed.
 \noindent
\vskip .5cm
\vskip .5cm
\vfill
\eject

\end{titlepage}

\setcounter{footnote}{0}

\small{
\tableofcontents}

\normalsize

\newpage
\renewcommand{\theequation}{{\rm\thesection.\arabic{equation}}}
%
\section{Introduction}
The AdS/CFT conjecture and its refinements \cite{Maldacena:1997re,Gubser:1998bc,Witten:1998qj} naturally lead to the application of holography to study non-conformal field theories at strong coupling. 

There are two well-established ways of constructing duals to confining QFTs. One of them uses wrapped branes--see for example \cite{Witten:1998zw,Maldacena:2000yy,Atiyah:2000zz,Edelstein:2001pu}. The second suitably deforms the solution of D3 branes on the conifold, introducing fractional branes and making contact with quiver field theories \cite{Klebanov:1998hh,Klebanov:2000nc,Klebanov:2000hb,Gubser:2004qj}. Both lines of work have been thoroughly studied and generalised in different ways. It is also possible to connect these two lines of study \cite{Maldacena:2009mw,Gaillard:2010qg}.

One issue with the systems discussed above is that while the IR of the QFT is under good control (thanks to the trustworthy string theory dual), the UV behaviour is less clean. In the wrapped brane systems we encounter a higher dimensional completion, which in some cases is non-field theoretical. On the other hand, the quiver system does not strictly speaking reach a conformal fixed point. For both lines of research, the number of degrees of freedom (from a 4d perspective) grows without bound at large energies. Aside from this, introducing the dynamics of  degrees of freedom transforming in the fundamental representation of the gauge group (quarks), is technically challenging. See \cite{Casero:2006pt,Nunez:2010sf,Benini:2006hh,Benini:2007gx,Bigazzi:2008qq,Bigazzi:2014qsa}. 

In this  paper, we present progress on the above shortcomings. We construct infinite families of string theory backgrounds dual to 4d SCFTs that flow to $(2+1)$-dimensional ${\cal N}=2$ IR-gapped QFTs. More details, both technical and conceptual, can be found in a companion paper \cite{Chatzis:2024kdu}.

Our new backgrounds are dual to families of 4d SCFTs deformed by VEVs and by compactification on a circle. The compactification resembles a twisting procedure. The low energy QFT is three dimensional, enjoys the existence of four supercharges, and is on a vacuum giving VEV to a current. These field theories can be Lagrangian or non-Lagrangian, single node or linear quiver type (with flavours). Field theoretically, the R-symmetry of the SCFT is used to preserve some amount of SUSY that would have otherwise been broken by the boundary conditions of fields on the compact $S^1$. In the case of ${\cal N}=4 $ SYM, this deformation has been carefully studied on the field theory side by Kumar and Stuardo in \cite{Kumar:2024pcz} and in the work of Cassani and Komargodski \cite{Cassani:2021fyv}. The main technical tools  for our work come from the papers \cite{Anabalon:2021tua,Anabalon:2022aig,Gauntlett:2007sm,bpt2017,afprt2015,Apruzzi:2015zna,plethora}.

\subsection{General idea and outline}\label{QFTapproach}
When a SUSY QFT is placed on a spacetime of the form $R^{1,d}\times S_\phi^1$, we need to specify boundary conditions for the fermions, scalars and gauge fields on $S_\phi^1$. Characteristically this breaks SUSY, as the scalars and gauge fields in the QFT obey periodic boundary conditions, whilst the fermions are anti-periodic.

It is possible to preserve some amount of SUSY by mixing the R-symmetry of the QFT with the part of the Lorentz group associated to translations on $S_\phi^1$.
See \cite{Kumar:2024pcz} for a nice explanation in the case of ${\cal N}=4$ SYM. The reader will also find instructive the viewpoint presented in \cite{Cassani:2021fyv}.

Indeed, switching on a constant background gauge field $\mathcal{A}=\mathcal{A}_\phi \mathrm{d}\phi$ for the R-symmetry changes the covariant derivative in such a way that allows massless fermions to exist. These pair up with the gauge field in a supermultiplet. Massive modes for fermions and scalars also pair up, preserving four supercharges. This leads to an ${\cal N}=2$ SYM theory in three dimensions (plus massive multiplets). The background gauge field is constant, but has non-zero holonomy, and hence dynamical effects.

At tree level, after reducing on $S_\phi^1$ and taking the low energy limit we have a pure
${\cal N}=2$ massless vector multiplet. This system presents a run-away potential and the vacuum is unstable. This is cured by observing that the KK-modes on the circle do not decouple \cite{Cassani:2021fyv}. In fact, these KK modes are non-symmetric with respect to the origin and integrating them out generates a Chern-Simons term. The theory flows to a  trivial confining vacuum. The level of the Chern-Simons term for a group SU($N$) is $N$, no matter what the original ${\cal N}=1$ theory in $d=4$ is. This result is also valid for quiver gauge theories and for non-lagrangian theories \cite{Cassani:2021fyv}.

The holographic dual to this particular form of twisting is described in a paper by Anabal\'on and Ross \cite{Anabalon:2021tua}. See also the papers \cite{Nunez:2023nnl, Anabalon:2024che, Anabalon:2024qhf, Fatemiabhari:2024aua,  
Nunez:2023xgl, Anabalon:2023lnk, Anabalon:2022aig} for other holographic studies in models using the same mechanism.

The idea of this work is to realise the mechanism of partial-SUSY preservation holographically for a variety of families of 4d SCFTs compactified on $S^1$. Our examples include single node CFTs,  and infinite families of circular and linear conformal quivers. We explicitly construct backgrounds that geometrically realise the mechanism above described. After that, we calculate various observable quantities for the QFT.

 The material in this work is divided in two main sections.

In Section \ref{section-geometry}, we discuss three new infinite families of backgrounds. These are of the form $\widehat{\text{AdS}}_5\times M_5$, where $\widehat{\text{AdS}}_5$ is a deformation of AdS$_5$ (and of the dual CFT$_4$) and $M_5$ encodes various characteristics of the CFT$_4$. The first of the new families is based on D3 branes on the tip of $Y^{p,q}$ cones \cite{Benvenuti:2004dy}.  The second new family is based on systems of D4-D6-NS5 branes that are dual in the UV to ${\cal N}=2$ SCFTs in four dimensions studied by Gaiotto and Maldacena \cite{Gaiotto:2009gz}. The third new family is based on a system of D6-D8-NS5 branes, which after compactification on a two-manifold leads to ${\cal N}=1$ SCFTs \cite{afpt2015,ct2015}. These SCFTs are then deformed as explained above. We present the backgrounds realising the deformation of the conformal point for all three families.  Our backgrounds are smooth, except at the location of the flavour branes (if any).

In Section \ref{QFT-section} we holographically calculate various observables for each dual QFT$_3$. In particular we identify signs of confinement using Wilson loops and Entanglement Entropy as our main tools. We also define a quantity that monotonically interpolates between  the number of degrees of freedom of a gapped 3d system and a 4d CFT. We briefly discuss some probes that display the presence of a Chern-Simons term of level given by the number of colour branes in the system, the presence of domain walls separating different vacua, and so on.

These interesting new backgrounds
can be used to obtain a better handle on the dynamics of QFTs displaying strongly coupled effects. Notably they allow for the holographic study of the effects of fundamental matter, with $N_f\sim N_c$ and SU(N$_f$)  global symmetry, introducing a novelty worth exploring in the future.

\section{Geometry: new families of backgrounds}\label{section-geometry}
In this section we present three new infinite families of backgrounds. Our presentation emphasises the common geometric origin in all these families. 
Indeed, even when the the dual QFT (studied in Section \ref{QFT-section}) is very different for each background, geometrically it is easy to see the common thread.

{The common thread mentioned above is due to a form of universality on the gravity side. Indeed, the solutions presented below are constructed by lifting a seed background in five-dimensional supergravity \cite{Anabalon:2021tua} to different Type II backgrounds. The details of the lifting procedures used are given in \cite{Chatzis:2024kdu}. In view of this, the reader might question the use of the adjective `new' for the solutions below, as all the novelty is in the `internal' space. We find that the while some of the Physics observables are universal (i.e. common to all backgrounds), some others are dependent on the lift.  
The universal sector alluded to above could be understood as the one predicted by Gauntlett and Varela in \cite{Gauntlett:2007ma} (see \cite{Chatzis:2024kdu} for a more detailed discussion).
}

We start by writing a background in the context of AdS$_5\times S^5$, which was already presented in \cite{Anabalon:2021tua}. We then change the internal space to $\mathrm{T}^{1,1}$ before presenting deformations of generic AdS$_5\times \mathrm{Y}^{p,q}$ backgrounds. After this we write other new infinite families of backgrounds in Type IIA and massive Type IIA, respectively.

\subsection{Deformed AdS$_5\times S^5$ and AdS$_5\times \mathrm{Y}^{p,q}$ backgrounds}
First consider the deformation of AdS$_5\times S^5$ studied in \cite{Anabalon:2021tua}. We give the metric, vielbein basis and five-form that solve the equations of motion of Type IIB. The metric is
 \begin{eqnarray} 
& &             \mathrm{d}s^2 _{10}  = \mathrm{d}s^2_{5}+ l^2 \biggl\{ \mathrm{d}\theta^2 + \sin^2\theta \mathrm{d}\varphi^2+ \sin^2\theta\sin^2\varphi \left( \mathrm{d}\varphi_1+ \frac{\mathcal{A}}{l} \right) ^2\nonumber\\
           & & + \sin^2\theta \cos^2\varphi \left( \mathrm{d}\varphi_2 + \frac{\mathcal{A}}{ l} \right) ^2 + \cos^2\theta \left( \mathrm{d}\varphi _3 + \frac{\mathcal{A}}{l} \right) ^2\biggr\},\label{metric-AdS5xS5}
\end{eqnarray} 
where the line element $\mathrm{d}s_5^2$ and one-form $\mathcal{A}$ are given by
\begin{eqnarray} 
           & &\mathrm{d}s^2_{5}=\frac{r^2}{l^2} (-\mathrm{d}t^2+\mathrm{d}x_1^2 + \mathrm{d}x_2^2 +  f(r)\mathrm{d}\phi^2) + \frac{l^2 ~\mathrm{d}r^2}{ r^2 f(r)} .\label{ds^2AR}\\
& &  f(r) = 1- \frac{\mu l^2}{r^4} - \frac{q^2l^2}{r^6},~~  \mathcal{A} =q \left( \frac{1}{r^2}- \frac{1}{r _{*}^2}\right)\mathrm{d}\phi \,.\label{ARgaugefield}
 \end{eqnarray} 
These expressions for $\mathrm{d}s^2_{5}$ and $\mathcal{A}$ (as well as its associated $\mathcal{F}=\mathrm{d} \mathcal{A}$) will appear in the backgrounds described in the sections to follow. In the remainder of this section, we choose the parameter $\mu=0$, as this implies the preservation of SUSY--- see \cite{Anabalon:2021tua} for details. The parameter $r_*= (q l)^{1/3}$ indicates the end of the space, for which $f(r_*)=0$. 

If we set the angle  $\phi$ to vary in $[0, \frac{2\pi l^2}{3r_*}]$ we avoid conical singularities and the background is smooth.
On the other hand, the range of the other angles is $\theta\in[0,\pi/2],\varphi\in[0,\pi/2]$,  $\varphi_1,\varphi_2,\varphi_3\in[0,2\pi]$. This space time defines a natural vielbein,
 \begin{equation}\label{vielbeinsAdS5xS5}
\begin{split}
       & e ^{1} = \frac{r}{l}\mathrm{d}t\, , \quad e ^{2}= \frac{r}{l}\mathrm{d}x    _1 \,\, , \,\, e ^{3}= \frac{r}{l}\mathrm{d}x_2\,\, , \,\, e ^{4}= \frac{l~ \mathrm{d}    r}{r \sqrt{f(r)}}\,\, , \,\, e ^{5}= \frac{r}{l}\sqrt{f(r)}\mathrm{d}\phi,\\
          & e^{6}=l \mathrm{d}\theta \,\, , \,\, e ^{7} = l\sin\theta \mathrm{d}\varphi \,\, , \,\, e ^{8} = l\sin\theta\sin\varphi \left( \mathrm{d}\varphi _1  + \frac{\mathcal{A}}{l} \right) , \\
           &e ^{9}= l\sin\theta\cos\varphi \left( \mathrm{d}\varphi _2 + \frac{\mathcal{A}}{l} \right) \,\, , \,\, e ^{10}=- l \cos\theta \left( \mathrm{d}\varphi_3 + \frac{\mathcal{A}}{l} \right). 
    \end{split}
   \end{equation}
%
It is also useful to introduce the quantities
\begin{eqnarray}
 & &
 \mu_1= \sin\theta \sin\varphi,~\mu_2= \sin\theta \cos\varphi, ~\mu_3= \cos\theta.\label{mui}
\end{eqnarray}
In terms of these we can write
\begin{eqnarray}
 & & F_5= ( 1+ {\star_{10} }) G_5, \qquad 
 G_5= -\frac{4}{l} e^{1}\wedge e^{2}\wedge e^{3}\wedge e^{4}\wedge e^{5} +  J_2\wedge {\star_5} \mathcal{F}, 
 \nonumber\\
 & & J_2= l^2 \sum_{i=1}^3 \mu_i \mathrm{d}\mu_i \wedge \left(\mathrm{d}\varphi_i +\frac{\mathcal{A}}{l}\right),
\qquad 
\mathcal{F}= \frac{2 q}{r^3}e^{5}\wedge e^4,
\qquad
\star_5\mathcal{F}= - \frac{2q}{r^3}e^{1}\wedge e^2\wedge e^3.\label{RR-S5}
\end{eqnarray}
Here, $\star_5$ is the Hodge star restricted on $\mathrm{ds}_5^2$.
We can  calculate the number of $\mathrm{D}3$ branes by computing the flux of the Ramond-Ramond field on the compact manifold $\Sigma_5[\theta,\varphi,\varphi_1,\varphi_2,\varphi_3]$. The quantisation condition for  $Dp$-branes reads
\begin{eqnarray}
& &	\int _{\Sigma _{8-p}}F _{8-p}= (2\pi) ^{7-p} g _{s} \alpha ^{\prime \frac{7-p}{2}}N _{\mathrm{D}p},~~\text{which for $p=3$ gives}\nonumber\\
& & 
	N _{\mathrm{D}3} = \frac{l^4}{4\pi g_s \alpha ^{\prime 2}}.\label{d3quantised}
\end{eqnarray}

%
Let us now present new families of solutions. In this family the five sphere is replaced by a $\mathrm{Y}^{p,q}$ manifold \cite{Gauntlett_2004,Martelli_2005}. The idea is very much the same as in the background of eqs.(\ref{metric-AdS5xS5})-(\ref{RR-S5}). We fiber a U(1)$_R$ isometry with the $\phi$-direction, which is compactified with a suitably chosen radius to avoid singularities. The functions in $\mathrm{d}s_5^2$ are chosen as in eq.(\ref{ARgaugefield}). A structure similar to that in eq.(\ref{RR-S5}) arises for the Ramond forms. As a warm-up we  discuss the case of $\mathrm{T}^{1,1}$.

\subsubsection{Deformed AdS$_5\times \mathrm{T}^{1,1}$ background}
 
The manifold $\mathrm{T}^{1,1}$ can be written as an $\mathrm{S}^1$ bundle over $\mathrm{S}^2\times\mathrm{S}^2$, parameterised by $(\theta_1,\phi_1,\theta_2,\phi_2)$,
\begin{equation}
\mathrm{d}s^2_{\mathrm{T}^{1,1}}=\frac{1}{6}\sum_{i=1}^2\left(\mathrm{d}\theta_i^2+\sin^2\theta_i\mathrm{d}\phi_i^2\right) + \frac{1}{9}\left(\mathrm{d}\psi +\sum_{i=1}^2 \cos\theta_i\mathrm{d}\phi_i\right)^2,
\end{equation}
where $\psi\in[0,4\pi]$. The Type IIB metric and vielbein are, 
\begin{eqnarray}
& &\mathrm{d}s^2_{10}=\mathrm{d}s^2_{5}+l^2\left[\frac{1}{6}\sum_{i=1}^2\left(\mathrm{d}\theta_i^2+\sin^2\theta_i\mathrm{d}\phi_i^2\right) + \frac{1}{9}\left(\mathrm{d}\psi + \sum_{i=1}^2\cos\theta_i\mathrm{d}\phi_i+\frac{3}{l}\mathcal{A}\right)^2\right].\label{AdS5xT11}\\
& & e ^{1} = \frac{r}{l}\mathrm{d}t\, , \quad e ^{2}= \frac{r}{l}\mathrm{d}x    _1 \,\, , \,\, e ^{3}= \frac{r}{l}\mathrm{d}x_2\,\, , \,\, e ^{4}= \frac{l~\mathrm{d}r}{r \sqrt{f(r)}}\,\, , \,\, e ^{5}= \frac{r}{l}\sqrt{f(r)}\mathrm{d}\phi,\nonumber\\
          & & e^{6}=\frac{l}{\sqrt{6}}\mathrm{d}\theta_1 \,\, , \,\, e ^{7} =\frac{l}{\sqrt{6}}\sin\theta_1\mathrm{d}\phi_1\,\, , \,\, e ^{8} =\frac{l}{\sqrt{6}}\mathrm{d}\theta_2  ,\nonumber\\
           & &e ^{9}=\frac{l}{\sqrt{6}}\sin\theta_2\mathrm{d}\phi_2 \,\, , \,\, e ^{10}=\frac{l}{3}\left(\mathrm{d}\psi+\cos\theta_1\mathrm{d}\phi_1+\cos\theta_2\mathrm{d}\phi_2+\frac{3}{l}\mathcal{A}\right). \nonumber
   \end{eqnarray}
The RR field strength $F_5$ is written in terms of the volume element of the five-spacetime $\mathrm{d}s_5^2$ and the field strength of the one form $\mathcal{A}= q \left( \frac{1}{r^2}- \frac{1}{r _{*}^2}\right)\mathrm{d}\phi $ as,
 \begin{eqnarray}
& &F_5=G_5+\star G_5,\label{F5T11}, \qquad G_5=\frac{4}{l} \mathrm{vol}_{5}-J\wedge\star_5\mathcal{F},\nonumber\\
& &\mathrm{vol}_{5}\!=\!e^1\wedge e^{2}\wedge e^{3}\wedge e^{4}\wedge e^{5}, \qquad 
J=-e^{6}\wedge e^{7}-e^{8}\wedge e^{9}.
\end{eqnarray}
The quantisation of charges works similarly to eq.(\ref{d3quantised}). We integrate over $\mathrm{T}^{1,1}$ and obtain
\begin{equation}
N_{D3}=\frac{4l^4}{27\pi g_s \alpha'^2}.
\end{equation}
\subsubsection{Deformed AdS$_5\times \mathrm{Y}^{p,q}$ backgrounds}
Following the previous section, we tackle the case of a general $\mathrm{Y}^{p,q}$ manifold. Things work in a similar fashion, though the metric is more involved, see 
\cite{Martelli_2005}. We find,
\begin{equation}\label{upliftYpq}
\begin{split}
\mathrm{d}s_{10}^2=\mathrm{d}s^2_{5}+ & l^2\left[\frac{1-y}{6}\left(\mathrm{d}\theta^2+\sin^2\theta\mathrm{d}\varphi^2\right)+\frac{1}{w(y)v(y)}\mathrm{d}y^2+\frac{w(y)v(y)}{36}\left(\mathrm{d}\beta+\cos\theta\mathrm{d}\varphi\right)^2\right.\\
   & \left.+\frac{1}{9}\left(\mathrm{d}\psi- \cos\theta\mathrm{d}\varphi+y\left(\mathrm{d}\beta+\cos\theta\mathrm{d}\varphi\right)+ \frac{3}{l}\mathcal{A}\right)^2\right],\\
    \end{split}
\end{equation}
with
\begin{equation}
    w(y)=\frac{2(a-y^2)}{1-y}\,\,,\,\,v(y)=\frac{a+2y^3-3y^2}{a-y^2},
\end{equation}
where $a$ is a parameter\footnote{In particular, for $0<a<1$ the base manifold $\mathrm{B}$ can have the topology of a product of two-spheres and the coordinate $y$ varies between the two smallest roots of $a-3y^2+2y^3=0$ which we call $y_1$ and $y_2$.}. The vielbein for this family of backgrounds is,
\begin{equation}
\begin{split}
       & e ^{1} = \frac{r}{l}\mathrm{d}t\, , \quad e ^{2}= \frac{r}{l}\mathrm{d}x    _1 \,\, , \,\, e ^{3}= \frac{r}{l}\mathrm{d}x_2\,\, , \,\, e ^{4}= \frac{l~ \mathrm{d}r}{r \sqrt{f(r)}}\,\, , \,\, e ^{5}= \frac{r \sqrt{ f(r)}}{l}\mathrm{d}\phi,\\
          & e^{6}=l \sqrt{\frac{1-y}{6}}\mathrm{d}\theta \,\, , \,\, e ^{7} =l \sqrt{\frac{1-y}{6}}\sin\theta\mathrm{d}\varphi\,\, , \,\, e ^{8} =\frac{l}{\sqrt{6}H(y)}\mathrm{d}y , \\
           &e ^{9}=l \frac{H(y)}{\sqrt{6}}(\mathrm{d}\beta+\cos\theta\mathrm{d}\varphi) \,\, , \,\, e ^{10}=\frac{l}{3}\left(\mathrm{d}\psi-\cos\theta\mathrm{d}\varphi+y(\mathrm{d}\beta+\cos\theta\mathrm{d}\varphi)+\frac{3}{l}\mathcal{A}\right).
    \end{split}
   \end{equation}
We have defined
$
    H(y)=\sqrt{\frac{wv}{6}}.
$
The K\"{a}hler form of the four dimensional base space spanned by $[\theta,\varphi,y,\beta]$ is
\begin{equation}
    J=e^6\wedge e^7+e^8\wedge e^9.
\end{equation}
The five-form flux is written in analogy with eq.\eqref{F5T11}:
\begin{equation}
\begin{split}
    F_5&=\frac{4}{l}\mathrm{vol}_{5}- \frac{4}{l} e^6\wedge e^7\wedge e^8\wedge e^9\wedge e^{10}+\frac{2q}{r^3}e^1\wedge e^2\wedge e^3\wedge J \label{ypqf5}\\
    &-\frac{2q}{r^3}e^4\wedge e^5\wedge(e^6\wedge e^7 + e^8\wedge e^9)\wedge e^{10}.
\end{split}
\end{equation}
The charge of D3 branes is quantised in analogy with eq.(\ref{d3quantised}).

We emphasize that all the backgrounds we have presented in this section solve the Einstein and Maxwell equations in Type IIB, together with the Bianchi identities. Additionally, all these solutions preserve four supercharges.

 Let us now present two qualitatively different families of backgrounds. The first is based on solutions in 11d supergravity written by Gaiotto and Maldacena \cite{Gaiotto:2009gz}, and the second family on massive Type IIA backgrounds written by Bah, Passias and Tomasiello in \cite{bpt2017}.

\subsection{Deformed Gaiotto-Maldacena backgrounds}
We study an new infinite family of backgrounds based on Gaiotto-Maldacena solutions \cite{Gaiotto:2009gz}. The idea is once again to preserve ${\cal N}=1$ SUSY (four supercharges) while deforming the original backgrounds by a fibration between the R-isometry--in this case SU(2)$\times$U(1)--and the compactified $\phi$-direction in $\mathrm{ds_5}^2$. 

Using 
the $\mu_i$ in eq.(\ref{mui}) and the one-form $\mathcal{A}$ in \eqref{ARgaugefield} we define
\begin{align}
D \mu_1
=\mathrm{d} \mu_1+2 \mu_2 \mathcal{A},&  \quad D \mu_2
=\mathrm{d} \mu_2-2\mu_1 \mathcal{A}, \quad  D \mu_3=\mathrm{d} \mu_3, \nonumber 
\\
\text{Vol}\tilde{S^2}&=\frac{1}{2}\epsilon^{ijk}\mu_i D\mu_j\wedge D\mu_k.
\label{GM-LLM2}
\end{align}
In terms of these forms and a single function $V(\sigma, \eta)$ and its derivatives, $\dot{V} \equiv \sigma\partial_\sigma V$ and  $V' \equiv \partial_\eta V$, we can write a configuration in eleven-dimensional supergravity which will automatically solve the equations of motion so long as the ``potential function'' $V(\sigma,\eta)$ satisfies the Laplace-like equation 
\begin{equation}
\sigma \partial_\sigma \left( \sigma \partial_\sigma V \right)+\sigma^2 \partial^2_\eta V=0. \label{laplace}
\end{equation}
The metric and four-form field strength read,
\begin{eqnarray}
& & \frac{\mathrm{d}s^2_{11}}{\kappa^{2/3}}= f_1\Big[4 \mathrm{d}s_5^2 + f_2 D\mu_i D\mu_i + f_3 \left(D\tilde{\chi}\right)^2 + f_4(\mathrm{d}\sigma^2+\mathrm{d}\eta^2)+ f_5\left(\mathrm{d} \tilde{\beta} + f_6 D\tilde{\chi}  \right)^2  \Big] ,\nonumber\\
& & \frac{G_4}{4\kappa}=  \, \mathrm{d}\left[ f_7 D \tilde{\chi}+f_8 \mathrm{d}\tilde{\beta}\right] \wedge \text{vol}\tilde{S}^2+  \,\mathrm{d}(\mu_3  \dot{V}) \wedge \star_5 \mathcal{F}
\nonumber\\
& &+2 (f_7 D \tilde{\chi}+f_8 \mathrm{d}\tilde{\beta}) \wedge \mathrm{d}\mu_3 \wedge \mathcal{F}- \left[ \mathrm{d}(\mu_3 \dot{V}) \wedge D\tilde{\chi}+\mathrm{d}(\mu_3 \eta) \wedge \mathrm{d}\tilde{\beta}\right]\wedge \mathcal{F}. \nonumber\\
& &   f_1=\bigg(\frac{\dot{V}\tilde{\Delta}}{2V''}\bigg)^{\frac{1}{3}},~~~~f_2 = \frac{2V''\dot{V}}{\tilde{\Delta}},~~~~f_3=\frac{4\sigma^2}{\Lambda},~~~~f_4 = \frac{2V''}{\dot{V}},\nonumber\\
& & f_5=\frac{2\Lambda V''}{\dot{V}\tilde{\Delta}},
    ~~~f_6=\frac{2\dot{V}\dot{V}'}{V''\Lambda},~~~
    f_7=-\frac{\dot{V}^2V''}{\tilde{\Delta}},~~f_8=\frac{1}{2}\bigg(\frac{\dot{V}\dot{V}'}{\tilde{\Delta}}-\eta\bigg),\nonumber
        \\ & &\tilde{\Delta} =\Lambda(V'')^2+(\dot{V}')^2,~~~~~~~~~\Lambda=\frac{2\dot{V}-\ddot{V}}{V''},~~~~~~~~~  D\tilde{\chi} = \mathrm{d}\tilde{\chi}+\mathcal{A}\label{def-GM}.
\end{eqnarray}
Suitable boundary conditions must be imposed on $V(\sigma,\eta)$. These boundary conditions encode the particular dual linear-quiver field theory, via a Rank function. This formalism is summarised in \cite{Nunez:2019gbg} and \cite{Macpherson:2024frt}.
It is in this sense that we have an infinite number of backgrounds in this family, each one associated with
a different 4d linear quiver ${\cal N}=2$ SCFT. In our setup these UV fixed points are deformed into a confining QFT by compactification on the $\phi$-circle.

Reducing to Type IIA along the $\beta$-circle in order to preserve SUSY (see \cite{Macpherson:2024frt} for details), we find the ten-dimensional string frame background,
\beq\label{eq:N=2} 
\begin{aligned}
\mathrm{d}s^2&= f_1^{\frac{3}{2}} f_5^{\frac{1}{2}}\bigg[4\mathrm{d}s^2_5+f_2 D\mu_iD\mu_i+f_4(\mathrm{d}\sigma^2+\mathrm{d}\eta^2)+f_3 (\mathrm{d}\tilde{\chi} +\mathcal{A})^2\bigg],\\[2mm]
e^{\frac{4}{3}\Phi}&=  f_1 f_5
,~~~~  H_3 = 4\k ~\mathrm{d}\left[f_8\wedge \text{vol}\tilde{S}^2-\eta  \mu_3\wedge \mathcal{F}\right],~~~~C_1=  f_6 D\tilde{\chi},~~~~,\\[2mm]
C_3&=4\k f_7 D\tilde{\chi}\wedge\text{vol}\tilde{S}^2+4\k ~\mu_3 \dot{V} \left(\star_5 \mathcal{F}-D\chi \wedge \mathcal{F}\right).\nn 
\end{aligned}
\eeq
We have used the relation $\mathrm{d}[\text{vol}\tilde{S}^2]=-2\mathrm{d}[\mu_3]\wedge \mathcal{F}$. Hence 
\beq\label{eq:F4} 
\begin{aligned}
F_4&=\mathrm{d} C_3 - H_3\wedge C_1 = 4\k ~\mathrm{d} \left[ f_7 D\tilde{\chi}\wedge\text{vol}\tilde{S}^2+\mu_3 \dot{V} \left(\star_5 \mathcal{F}-D\tilde{\chi} \wedge \mathrm{d} \mathcal{A}\right)\right]-H_3\wedge C_1.\nn 
\end{aligned}
\eeq
We have checked that the equations of motion of this background are satisfied, once eq.(\ref{laplace}) is imposed. Following the study in \cite{Macpherson:2024frt} we construct the Page fluxes and find  quantised number of D4, D6 and NS five branes. We postpone the careful discussion of the charge quantisation to the more detailed work \cite{Chatzis:2024kdu}.

\subsection{Deformed D6-D8-NS5 AdS$_5$ backgrounds}
In this section we consider deformed holographic duals of an infinite family of four dimensional non-Lagrangian CFTs. The  backgrounds dual to the CFT fixed points are obtained by a twisted compactification over a hyperbolic space of  a system of D6, D8 and NS5 branes \cite{afprt2015,Apruzzi:2015zna,bpt2017}. A further twisted compactification of the $\phi$-direction of the CFT$_4$ leads to the confining 3d system.

The background metric in massive type IIA is written in terms of a function $\alpha(z)$ satisfying
\begin{equation}
  \dddot{\alpha}(z)=-162\pi^3 F_0.\label{CT-eq}
  \end{equation}
Here the RR field $F_0$ is the `mass' of massive IIA, which is allowed to be piece-wise continuous and constant. This third order ordinary differential equation needs to be complemented by initial conditions. The conditions on $\ddot{\alpha}(z)$ encode the dual UV-CFT. See \cite{Merrikin:2022yho,Nunez:2018ags} for a clear account.
The background metric and the dilaton read,
\begin{eqnarray}
   & &  \mathrm{d}s_{10}^2=18\pi\sqrt{-\frac{\alpha}{6 \Ddot{\alpha}}}\left[\mathrm{d}s_5^2+\frac{1}{3} \mathrm{d}s_\Sigma^2-\frac{\Ddot{\alpha}}{6\alpha}\mathrm{d}z^2-\frac{\alpha \Ddot{\alpha}}{6 \dot{\alpha}^2-9 \alpha \Ddot{\alpha}} \left(\mathrm{d}\theta^2+\sin^2 \theta \mathcal{D}\psi^2\right)
    \right],\nonumber\\
    & &e^{-4\Phi}= \frac{1}{2^5 3^{17}\pi^{10}}\left( -\frac{\ddot{\alpha}}{\alpha}\right)^3 \left( 2\dot{\alpha}^2-3 \alpha \ddot{\alpha}\right)^2,~~~{\cal D}\psi= \mathrm{d}\psi - 3\mathcal{A} - A_\Sigma \label{metric-dil-BPT}\\
    & & \mathcal{A}=q\left(\frac{1}{r^2}-\frac{1}{r_*^2} \right)  \mathrm{d}\phi,~~A_\Sigma= \frac{2(v_1 ~\mathrm{d}v_2- v_2 ~ \mathrm{d}v_1)}{1-v_1^2-v_2^2},~~~  \mathrm{d}s_\Sigma^2= 4\frac{(\mathrm{d}v_1^2+\mathrm{d}v_2^2)}{(1- v_1^2-v_2^2)^2}.\nonumber
\end{eqnarray}
The Ramond and Neveu-Schwarz potentials and their associated field strengths can be compactly written as
\begin{eqnarray}
& &  B_2 =\frac{1}{3} \xi \wedge \mathcal{D}\psi, 
       \qquad 
    C_1 = \frac{\Ddot{\alpha}}{162\pi^2}\cos \theta\, \mathcal{D}\psi, \qquad 
    C_3 = \frac{\dot{\alpha}}{162\pi} \mathcal{D}\psi \wedge \text{vol}_\Sigma, \nonumber\\
& &
H_3=\mathrm{d}B_2, 
    \qquad 
   F_2 =F_0 B_2 + \mathrm{d}C_1,\label{RR-NS-BPT}
\\
& & F_4 =\left(\mathrm{d}C_3 +B_2  \wedge F_2 -\frac{1}{2} F_0\, B_2  \wedge B_2\right) -\frac{\Ddot{\alpha}}{18\pi}\mathrm{d}z  \wedge \left(\star_5 \mathcal{F}-\frac{1}{3}\mathcal{F} \wedge \mathcal{D} \psi\right) -\frac{\dot{\alpha}}{54\pi}\mathcal{F}\wedge \text{vol}_\Sigma  , \nonumber
\end{eqnarray}
in terms of a one form $\xi$, the two form $\mathcal{F}=\mathrm{d}\mathcal{A}$, and its Hodge dual in five dimensions,
\begin{equation}
      \xi=3\pi \left( \cos\theta  \mathrm{d}z-\frac{2\alpha \dot{\alpha}}{2\dot{\alpha}^2-3\alpha \Ddot{\alpha}} \sin \theta  \mathrm{d}\theta \right),~~\mathcal{F}= -\frac{2 q}{r^3}  \mathrm{d}r\wedge  \mathrm{d}\phi,~~\star_5 \mathcal{F}=-2q~  \mathrm{d}t\wedge  \mathrm{d}x_1\wedge  \mathrm{d}x_2.
\end{equation}
We have checked that all equations of motion (Einstein, Maxwell and Bianchi)
are solved once eq.(\ref{CT-eq}) is imposed.

Let us close this section with some general remarks. As in the previous section, all the backgrounds presented contain the same $\mathrm{d}s_5^2$ given in eq.(\ref{ARgaugefield}). All the backgrounds in eqs.(\ref{metric-AdS5xS5}), (\ref{AdS5xT11}), (\ref{def-GM}), (\ref{metric-dil-BPT}) are smooth once the period of the $\phi$-coordinate is chosen as explained below eq.(\ref{ARgaugefield}). They are all asymptotically AdS$_5$ with compact $\phi$-direction, cross an internal manifold, and the deformation away from AdS$_5$ is the same in each case--- namely, the compact $\phi$-direction is warped by the function $f(r)$. A U(1)$_R$ inside the internal manifold is fibered over the $\phi$-direction (and, if needed to preserve SUSY, over additional compact manifolds). For $\mu=0$, and thus $f(r)=1-\frac{q^2 l^2}{r^6}$, the backgrounds preserve four Poincare supercharges.
{These common threads in the geometry reflect the fact that all the backgrounds presented can be obtained as the lift to Type II supergravities of a common seed-solution in 5d $U(1)$ gauged supergravity: the background and gauge field in eqs.(\ref{ds^2AR})-(\ref{ARgaugefield}). As expected, this has consequences in the dual QFTs. In these, there are universal observables. There are also some observables that depend explicitly on the particular member of the infinite family of backgrounds presented.}

We move now to explore some holographic aspects of these QFTs as they flow towards 3d gapped theories in the IR.


\section{Field theory and observables}\label{QFT-section}
In this section we discuss holographic observables for the QFTs dual to the backgrounds in Section \ref{section-geometry}. We observe certain universal
behaviours across all the families of solutions, reflecting common features of their dual field theories. Some other physically interesting observables are presented which distinguish the QFTs. We start with a brief discussion of the dual QFT for each family of backgrounds. Then we move into calculating observables.

\subsection{Comments on the dual QFTs}
Expanding on the discussion in Section \ref{QFTapproach}, we write here some comments on the QFT duals in view of the backgrounds in Section \ref{section-geometry}.

Consider ${\cal N}=4$ SYM compactified on a circle of fixed size $L_\phi= \frac{2\pi l^2}{3 r_*}$. The mixing between the R-symmetry and the compact direction is holographically represented by the one-form $\mathcal{A}$ in eq.(\ref{ARgaugefield}). Were it not for this mixing, the background would break SUSY due to the boundary conditions on the $\phi$-circle. 
Note that for the coordinate $r\to\infty$ we have a constant one form $\mathcal{A}$ with a non-trivial holonomy $\oint_\phi \mathcal{A} = \frac{q}{r_*^2}L_\phi= \frac{2\pi}{3} l$. This is the manifestation in gravity of the twisting mechanism  described in Section \ref{QFTapproach}.

The presence of the one-form, together with the function $f(r)$ in eq.(\ref{ARgaugefield}), break conformality due to the scale introduced by $L_\phi$. Flowing to low energies leads to a $(2+1)$ dimensional QFT with strongly coupled IR dynamics.  This mechanism occurs in all the examples in this work. The low energy QFT is expected to confine, present discrete vacua, and to be described by a Chern-Simons TQFT with level $N$ (the number of colour branes) below the gap, see \cite{Cassani:2021fyv}, \cite{Kumar:2024pcz}.

The first background presented here implements this `twisted compactification' for ${\cal N}=4$ SYM, see eqs.(\ref{metric-AdS5xS5})-(\ref{RR-S5}). 
The solution given in eqs.(\ref{AdS5xT11})-(\ref{F5T11}) is holographically dual to the same compactification of the Klebanov-Witten CFT. Notice that in this case the CFT does not admit a weakly coupled regime. The holographic description of quiver field theories realised on D3 branes on the tip of a cone over a generic Sasaki-Einstein space (after twisted compactification) is presented by in eqs.(\ref{upliftYpq})-(\ref{ypqf5}).

This twisted compactification procedure for ${\cal N}=2$ 4d linear quiver theories is subtle. In fact, we need to twist the $\phi$-symmetry  with a U(1) that is inside SU(2)$_R\times$ U(1)$_R$. This is represented by the one form $\mathcal{A}$ in eq.(\ref{ARgaugefield}). Note that we are performing the same procedure for each gauge group in the linear quiver. Each of these linear quivers are characterised by a rank function used in the resolution of eq.(\ref{laplace})--- see \cite{Nunez:2019gbg,Macpherson:2024frt} for the explicit holographic solution.

 The example given in eqs.(\ref{CT-eq})-(\ref{RR-NS-BPT}) describes the twisted compactification of  ${\cal N}=1$ non-Lagrangian CFTs---one for each function $\alpha(z)$--- as explained for example in \cite{Nunez:2018ags,Filippas:2019puw}. In spite of the absence of a Lagrangian, we can perform (holographically) the twisted compactification and study the IR strong dynamics of the effective ${\cal N}=2$, $(2+1)$-dimensional QFT. Note that in all of our examples the R-symmetry is non-anomalous.

Applying holographic renormalisation one learns for all of our examples that a current $J_\phi$  is obtaining a VEV, $\langle J_\phi \rangle ~ \sim ~ q$.
Hence we are studying families of 3d QFTs with four supercharges in a vacuum that gives a VEV to a global current.  
Had we set the parameter $\mu\neq 0$ in $f(r)$ in eq.(\ref{ARgaugefield}) we would have broken SUSY due to an (anistropic) VEV for $T_{\mu\nu}$. See \cite{Anabalon:2021tua,Anabalon:2022aig,Fatemiabhari:2024aua}.

In what follows, we calculate various observables using the holographic backgrounds. One feature that becomes apparent is that various observable quantities can be written as a contribution coming from the flow, times another  contribution coming from the 4d UV SCFT. This is particularly clear in the calculations of Entanglement Entropy or flow central charge presented below. This phenomenon is the QFT expression of the proposal by Gauntlett and Varela \cite{Gauntlett:2007ma}, indicating that if one considers fields in the supercurrent multiplet of the SCFT, the ten dimensional supergravity background can be reduced to minimal AdS$_5$ gauged supergravity, which is the 'universal' part associated with the flow above mentioned. Note that all of our backgrounds can be obtained as lifts of a seed-solution in this gauged supergravity. This perspective is discussed in detail in the companion work \cite{Chatzis:2024kdu}.

\subsection{Observables}
We discuss here various observables calculated using the backgrounds in Section \ref{section-geometry}. We briefly sketch the calculation in the various backgrounds, stressing when a universal result is obtained. 

\subsubsection{Wilson loops}
First we compute the Wilson loop for the backgrounds in Section \ref{section-geometry}. We use the standard holographic prescription  of embedding a string with its endpoints fixed at $x=\pm L/2$ and $r=\infty$ (see e.g. \cite{Sonnenschein:1999if}). Fixing the worldsheet coordinates on the string as $\tau=t$, $\sigma=x_1\equiv x$, with $r=r(x)$ and all other coordinates are taken to be constant, the induced metric on the string and the associated Nambu-Goto action are
\begin{eqnarray}
& &  \mathrm{d}s^2 _{\mathrm{ind}}= - \frac{r^2}{l^2} \mathrm{d}t^2 + \left( \frac{r^2}{l^2}+ \frac{l^2}{r^2 f(r)}r ^{\prime 2} \right) \mathrm{d}x^2\nonumber,\\
& & 	S _{\mathrm{NG}} = \frac{\mathcal{T}}{2\pi\alpha ^{\prime}}\int _{-L/2} ^{L/2}\mathrm{d}x \sqrt{\frac{r^4}{l^4} + \frac{r ^{\prime 2}}{f(r)}}= \frac{\mathcal{T}}{2\pi\alpha ^{\prime}}\int _{-L/2} ^{L/2}\mathrm{d}x \sqrt{F(r)^2 + G(r)^2 r ^{\prime 2}} ,\nonumber\\
& & 	F(r) = \frac{r^2}{l^2}\,\, , \,\, G(r) = \frac{1}{\sqrt{f(r)}}.\label{functions-wilson}
\end{eqnarray}
It is important to note that given this embedding of the F1 string, the results in eq.(\ref{functions-wilson}) are {\it universal} for all of the backgrounds in Section \ref{section-geometry}. This is because these solutions all contain the same five-dimensional metric on the directions probed by the string, up to overall warp factors depending on the internal coordinates. Thus for all coordinates except $t,r,x_1$ constant, the various backgrounds differ only by constant factors re-scaling the induced metric.   Consequently the results for Wilson loops discussed below are valid for any of the dual QFTs.

The range of the radial coordinate is 
$[r_*, \infty)$, where $r_*$ is the largest root of $f(r_*)=0$. We usually consider the parameter $\mu=0$ in eq.(\ref{ARgaugefield}), hence $r_*^6=q^2l^2$.
The value of $F(r_*)=\frac{r_*^2}{l^2}$ is related to the effective tension of the chromoelectric string $T _{\mathrm{eff}}$ (see \cite{Kol:2014nqa,Faedo:2013ota,Faedo:2014naa,Kinar:1998vq}). This hints at the confining behaviour of these systems.  A more detailed analysis ascertaining the presence of confinement follows\footnote{To be generic in the analysis, we keep the parameter $\mu$ (for the SUSY case of our interest, $\mu=0$). }.

There is an approximate expression for the separation of the quark pair as a function of the turning point of the string probe $r_0$. This was derived in \cite{Kol:2014nqa}, \cite{Faedo:2014naa}. The approximate length reads in our case,
\begin{equation}\label{Lapp-Wilson}
	\hat{L}(r_0) = \left.\frac{\pi G(r)}{F^{\prime}(r)}\right| _{r_0}= \frac{\pi l^2 r_0^2}{2 \sqrt{r_0^6-\mu l^2 r_0^2 - q^2l^2}}.
\end{equation}
When $r_0\to r_{*}$, i.e. the turning point approaches the end of the spacetime, the approximate length diverges, indicating that it is possible to arbitrarily separate the quark pair. Following \cite{Nunez_2010} we introduce an effective potential,
\begin{equation}
	V _{\mathrm{eff}}(r) = \frac{F(r)}{F(r_0)G(r)}\sqrt{F(r)^2-F(r_0)^2}=\frac{\sqrt{r^4-r_0^4}}{r_0^2 l^2 r} \sqrt{r^6-l^2 (\mu r^2 + q^2)}.
\end{equation}
This approaches zero as $r\to r_0$ and diverges $\sim r^4$ for $r\to \infty$, satisfying the conditions needed for the proper definition of the Wilson loop and confinement--- see \cite{Nunez_2010}. These are more indications of the confining behaviour. The signature of the confining character of the backgrounds comes from studying 
 the separation and energy of the quark-anti-quark pair. These can be read from  \cite{Nunez_2010} to be,
\begin{eqnarray}
& &	L _{QQ}(r_0) =2 \int _{r_0}^{\infty}\frac{\mathrm{d}r}{V_{\mathrm{eff}}(r)}= 2 l^2 r_0^2 \int _{r_0} ^{\infty} \frac{\mathrm{d}r~ r}{\sqrt{\left( r^4-r_0^4 \right) \left[ r^6-l^2(q^2+r^2\mu) \right] }},\label{L-Wilson}\\
& & 	E _{QQ}(r_0) = \frac{r_0^2}{l^2}L _{QQ}(r_0) + \frac{2}{l}\int _{r_0}^{\infty} \mathrm{d}r\frac{\sqrt{r^4-r_0^4}}{r^2 \sqrt{f(r)}} - 2 \int _{r _{*}} ^{\infty} \frac{\mathrm{d}r ~ 
}{\sqrt{f(r)}}.\label{energy-Wilson}
\end{eqnarray}
Whilst these integrals are not analytically calculable in general, we can numerically find $L_{QQ}(r_0)$,
$E_{QQ}(r_0)$ and then parametrically plot $E_{QQ}(L_{QQ})$. The results are displayed in 
Figure \ref{PlotsforL(r0)E(r0)E(L)-Wilson-AdS5xS5}. We see that the scaling of the energy with the separation of the quark-anti-quark-pair is Coulomb-like for small separations, as required by conformality, and becomes linear for large separations, indicating confinement.

A careful study of the stability of the string embedding used here should follow the analysis of
\cite{Chatzis:2024dlt}. A short-cut is to
calculate the derivative of the length function with respect to the turning point of the $\mathrm{U}$-shaped string profile \cite{Faedo:2013ota,Faedo:2014naa,Nunez:2023nnl}. The approximate expression \eqref{Lapp-Wilson} yields,
\begin{equation}
    Z(r_0):=\frac{\mathrm{d}\hat{L}(r_0)}{\mathrm{d}r_0}= - \frac{l^4\pi }{4 r_0^3 ~f^{3/2}(r_0)}\left( \frac{4q^2}{r_0^5}+ \frac{2r_0}{l^2} + \frac{2\mu}{r_0^3}\right) <0.
\end{equation}
Being negative, this quantity indicates that the embedding is stable. {Notice that in this work, we choose the parameter $\mu\geq 0$.}

{A natural question regards the SUSY preservation of the Wilson loop studied above (we consider the case $\mu=0$). Given that our embedding does not excite the scalar associated with the R-symmetry of the QFT, it is likely that our probe F1 is not SUSY. A nice way of identifying supersymmetry-preserving embeddings (probably requiring a combination of F1 and D1 branes) would be to calculate the pure spinors characterising our backgrounds, for the parameter $\mu=0$. Knowing the pure spinors would allow to calculate SUSY probes in general. We expect that in the case of quiver field theories, Wilson loops in higher representations could be SUSY, as discussed for the CFT points in the papers \cite{Uhlemann:2020bek}, \cite{Fatemiabhari:2022kpv}.}

\begin{figure}[t]
  \centering
  \includegraphics[width=0.4\textwidth]{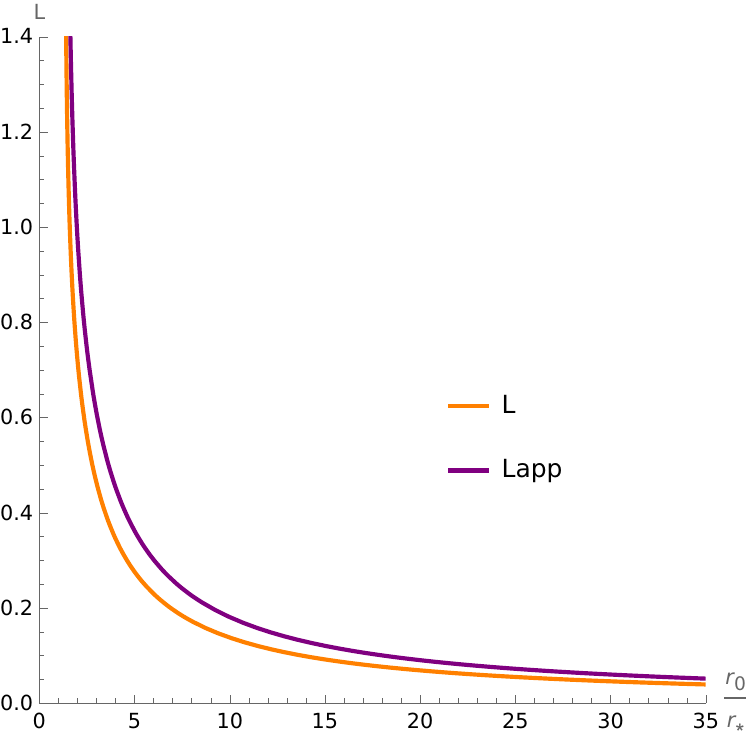}
 \hspace{0.3cm} 
  \centering
      \includegraphics[width=0.4\linewidth]{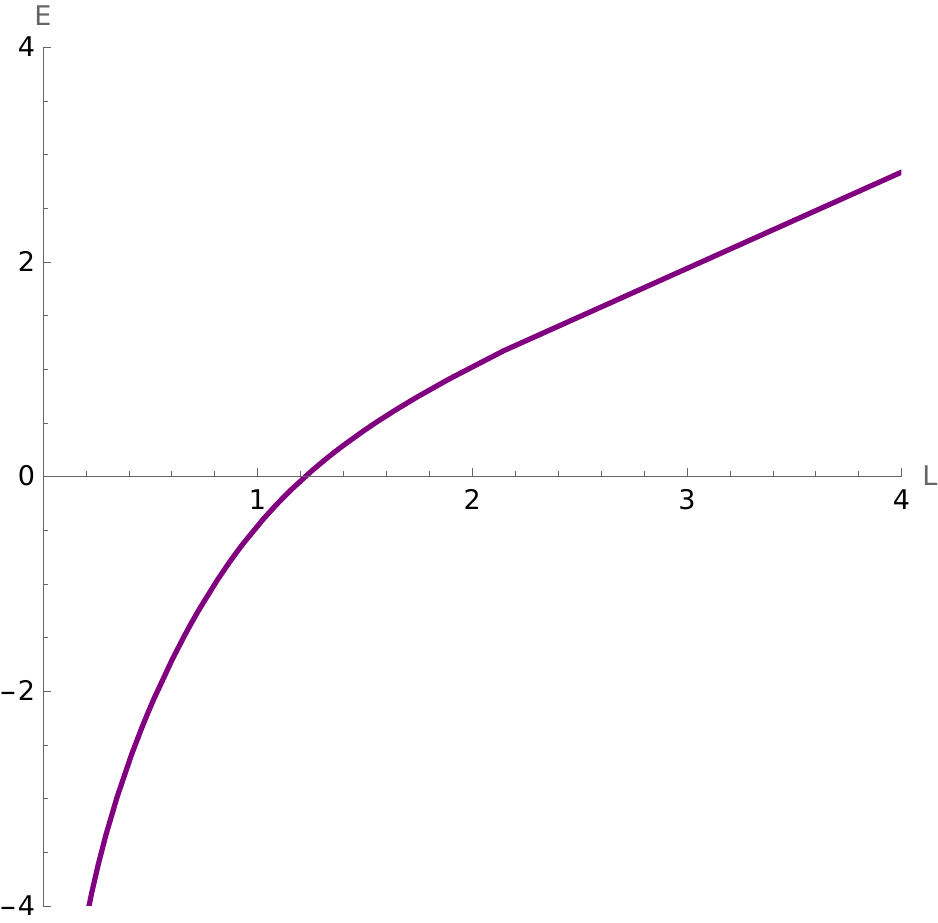}
\caption{On the left we plot the length of separation of the quark-anti-quark system \eqref{L-Wilson} as a function of the turning point of the string $r_0$. The parameters are fixed to $l=\mu=q=1$. On the right is the energy \eqref{energy-Wilson} with respect to the length of the separation of the pair, which interpolates between a Coulomb-like behaviour dictated by conformality and a linear one for large values of $L$, signaling confinement.}
\label{PlotsforL(r0)E(r0)E(L)-Wilson-AdS5xS5}
\end{figure}

We emphasise two points. First is the universal character of these results across the backgrounds we have presented. This is obvious from holography and far from obvious from a QFT perspective. In fact, the confining character of the QFT can only be ascertained via holography. Second, these results assumed a particular F1 string embedding. It would be interesting to study more generic embeddings probing the internal manifold M$_5$. This might reveal new physical phenomena.

Let us now study another observable, the Entanglement Entropy (EE) on a strip.

\subsubsection{Entanglement Entropy on the strip}
To calculate the EE on a strip we follow the usual prescription of Ryu and Takayanagi \cite{Ryu:2006bv}, with the requisite modifications for backgrounds with other NS fields excited \cite{Klebanov:2007ws,Kol:2014nqa}. First, we define an eight-manifold and calculate the determinant of the induced metric on this surface, weighted by a power of the dilaton. Below we write the eight-manifold and the induced metric for each of the backgrounds we presented in Section \ref{section-geometry}. 
\begin{itemize}
    \item For the case of the deformed AdS$_5 \times S^5$ background in eq.(\ref{metric-AdS5xS5})
\begin{eqnarray}
& & \Sigma_{8,S^5}=[x_1,x_2,\phi, \theta,\varphi, \varphi_1,\varphi_2,\varphi_3],~~~~~~ r(x_1),\nonumber\\
& & \mathrm{d}s_{8, S^5}^2= \frac{r^2}{l^2}\left( \mathrm{d}x_1^2\left(1+ \frac{l^4 r'^2}{r^4 f(r)}\right)+ \mathrm{d}x_2^2+ f(r)\mathrm{d}\phi^2 \right) +  l^2 \biggl\{ \mathrm{d}\theta^2 + \sin^2\theta \mathrm{d}\varphi^2+ \nonumber\\
& &\sin^2\theta\sin^2\varphi \left( \mathrm{d}\varphi_1+ \frac{\mathcal{A}}{ l} \right) ^2  + \sin^2\theta \cos^2\varphi \left( \mathrm{d}\varphi_2 + \frac{\mathcal{A}}{ l} \right) ^2 + \cos^2\theta \left( \mathrm{d}\varphi _3 + \frac{\mathcal{A}}{ l} \right) ^2\biggr\}.\nonumber \label{metric8S5} 
\end{eqnarray}
    \item For the deformed AdS$_5 \times \mathrm{Y}^{p,q}$ backgrounds in eq.(\ref{upliftYpq}), we find
\begin{eqnarray}
& & \Sigma_{8,\mathrm{Y}^{p,q}}=[x_1,x_2,\phi, \theta,\varphi, y,\beta,\psi],~~~~~~ r(x_1),\nonumber\\
& & \mathrm{d}s_{8, \mathrm{Y}^{p,q}}^2= \frac{r^2}{l^2}\left( \mathrm{d}x_1^2\left(1+ \frac{l^4 r'^2}{r^4 f(r)}\right)+ \mathrm{d}x_2^2+ f(r)\mathrm{d}\phi^2 \right) +l^2
\biggl\{    \frac{1-y}{6}\left(\mathrm{d}\theta^2+\sin^2\theta\mathrm{d}\varphi^2\right)+\nonumber\\
& &\frac{\mathrm{d}y^2}{w(y)v(y)}+ \frac{w(y)v(y)}{36}\left(\mathrm{d}\beta+\cos\theta\mathrm{d}\varphi\right)^2  +\frac{1}{9}\left(\mathrm{d}\psi- \cos\theta\mathrm{d}\varphi+y\left(\mathrm{d}\beta+\cos\theta\mathrm{d}\varphi\right)+ \frac{{3}\mathcal{A}}{l}\right)^2 \biggr\}. \nonumber
\label{metric8ypq}
\end{eqnarray}
    \item For the deformation on the Gaiotto-Maldacena backgrounds in eq.(\ref{eq:N=2}) we have,
\begin{eqnarray}
& &  \Sigma_{8,GM}=[x_1,x_2,\phi, \theta,\varphi, \tilde{\chi},\sigma,\eta],~~~~~~ r(x_1),\nonumber\\
& & ds_{8, GM}^2= \left( f_1^3 f_5\right)^{\frac{1}{2}}\biggl\{  \frac{4r^2}{l^2}\left( \mathrm{d}x_1^2\left(1+ \frac{l^4 r'^2}{r^4 f(r)}\right)+ \mathrm{d}x_2^2+ f(r)\mathrm{d}\phi^2 \right) +
\nonumber\\
& &f_2 D\mu_iD\mu_i+f_4(\mathrm{d}\sigma^2+\mathrm{d}\eta^2)+ f_3 (\mathrm{d}\tilde{\chi} +\mathcal{A})^2\biggr\}.\nonumber
\label{metric8GM}
\end{eqnarray}
This could also be done by calculating the area of a nine-manifold using the metric in eq.(\ref{def-GM}).
\item Finally, for the deformation of the Bah-Passias-Tomasiello family of backgrounds, we have
\begin{eqnarray}
& & BPT:~ \Sigma_{8,BPT}=[x_1,x_2,\phi, \theta,\psi, v_1,v_2, z],~~~~~~ r(x_1),\nonumber\\
& & \mathrm{d}s_{8, BPT}^2= 18\pi\sqrt{-\frac{\alpha}{6\ddot{\alpha}}}\biggl\{  \frac{r^2}{l^2}\left( \mathrm{d}x_1^2\left(1+ \frac{l^4 r'^2}{r^4 f(r)} \right)+ \mathrm{d}x_2^2+ f(r)\mathrm{d}\phi^2 \right) 
+\frac{1}{3} \mathrm{d}s_\Sigma^2-\frac{\Ddot{\alpha}}{6\alpha}\mathrm{d}z^2\nonumber\\
& & -\frac{\alpha \Ddot{\alpha}}{6 \dot{\alpha}^2-9 \alpha \Ddot{\alpha}} \left(\mathrm{d}\theta^2+\sin^2 \theta \mathcal{D}\psi^2\right)
\biggr\}. \nonumber
\label{metric8BPT}
\end{eqnarray}
\end{itemize}
To find the EE we calculate 
\cite{Klebanov:2007ws,Kol:2014nqa}
\begin{equation}
S_{\text{EE}}=\frac{1}{4 G_N}\int \mathrm{d}^8x \sqrt{e^{-4\Phi}\det[g_{{}_{\Sigma_8}}]}.  \label{EEgeneral} 
\end{equation}
The result in each case is of the form
\begin{equation}
    S_{\text{EE}}=\frac{{\cal N}_i}{4 G_N} \int_{-L/2}^{L/2} \mathrm{d}x~\sqrt{\frac{r^6}{l^6} f(r) \left(1+ \frac{l^4 r'^2}{r^4 f(r)} \right)}, \label{EEdiego}
\end{equation}
where we have introduced the label $i \in \{S^5, \mathrm{Y}^{p,q}, GM, BPT\}$, and 
\begin{eqnarray}
& & {\cal N}_{S^5}=  L_{x_2}l^5 \int_{0}^{\pi/2} \sin^3\theta \cos\theta ~\mathrm{d}\theta~\int_0^{\pi/2}\sin\varphi \cos\varphi ~\mathrm{d}\varphi \int_0^{2\pi}\mathrm{d}\varphi_1~\mathrm{d}\varphi_2~\mathrm{d}\varphi_3 \int_0^{L_\phi} \mathrm{d}\phi= l^5 L_{x_2} \pi^3 L_\phi ,\nonumber\\
& & {\cal N}_{\mathrm{Y}^{p,q}}= L_{x_2} l^5 \int_{y_1}^{y_2} \mathrm{d}y (1-y)\int_0^\pi \sin\theta \mathrm{d}\theta\int_0^{2\pi}\mathrm{d}\varphi \int_0^{2\pi} \mathrm{d}\beta \int_0^{4\pi}\mathrm{d}\psi  \int_0^{L_\phi} \mathrm{d}\phi=l^5 L_{x_2}\text{Vol}_{\mathrm{Y}^{p,q}} L_\phi,\nonumber\\
& & {\cal N}_{GM}= 256\pi^2 L_{x_2} L_\phi \int_0^\infty \mathrm{d}\sigma \int_0^P \mathrm{d}\eta~ \sigma \dot{V} {V''}\  ,\nonumber\\
& & {\cal N}_{BPT}= \frac{2}{243} L_{x_2} L_\phi \text{Vol}_{\Sigma}\int_0^P \mathrm{d}z(-\alpha\ddot{\alpha})  .\nonumber
\end{eqnarray}
\begin{figure}
  \centering
  \includegraphics[width=0.4\textwidth]{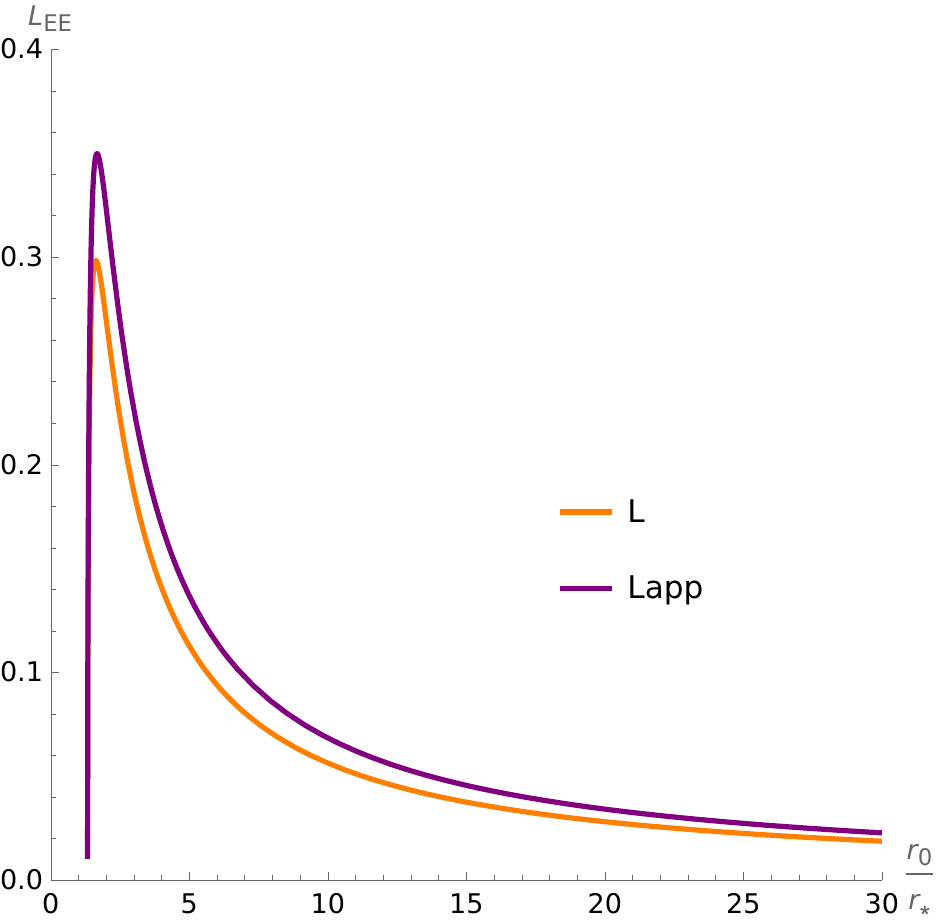}
 \hspace{0.4cm} \centering
  \includegraphics[width=0.4\textwidth]{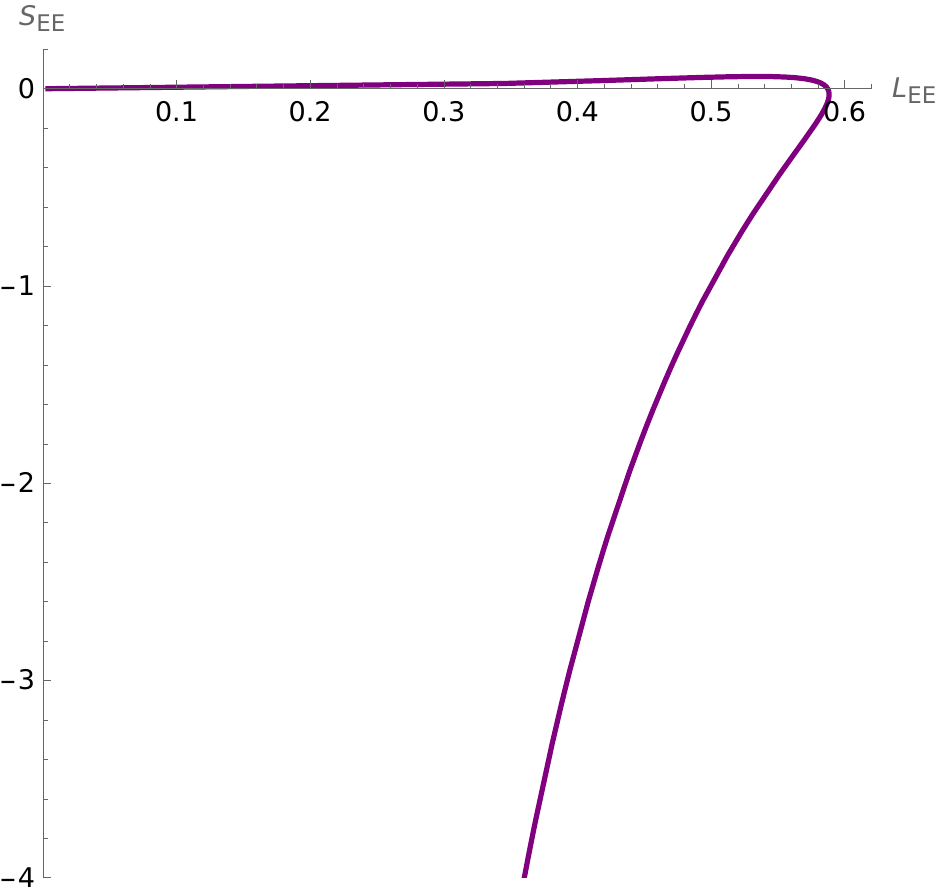}
\caption{On the left, we plot the length function and its approximate expression in terms of $r_0$ (the turning point). On the right is the EE as a function of the length. }
\label{EEfigure}
\end{figure}
Let us discuss these expressions. The result for the EE can be broken into two parts. One comes from the flow away from the CFT, represented by the integral in eq.(\ref{EEdiego})--note that for $f(r)=1$ we obtain the CFT result. This integral in eq.(\ref{EEdiego}) is universal across our backgrounds and is discussed in more detail below. The second part, the coefficient ${\cal N}_i$, is characteristic of the particular background (or dual CFT). It encodes information coming from the far UV of the QFT, at the conformal point. In fact, ${\cal N}_{S^5}$ and ${\cal N}_{\mathrm{Y}^{p,q}}$ are directly related to the volume of the internal manifold at the UV fixed point and give the central charge of the dual CFT. Similarly, ${\cal N}_{GM}$ and ${\cal N}_{BPT}$ are directly related to the free energies, or  central charges, of the CFT in the UV. One can compare the results for ${\cal N}_{GM}$ and ${\cal N}_{BPT}$ with eq.(2.37) in the paper \cite{Nunez:2019gbg} and eq.(2.21) in \cite{Filippas:2019puw}, or eq.(2.14) in \cite{Nunez:2018ags}, to see this. This field theory observable is decomposed as a part coming from the deformed AdS$_5$ (the flow) and another coming from the UV-CFT$_4$ (the coefficient ${\cal N}_i$). This is how the Gauntlett-Varela proposal \cite{Gauntlett:2007ma} expresses itself in this QFT observable.

Let us come back to the expression  for the EE (\ref{EEdiego}), which is calculated using an integral of the form
\begin{equation}
 \int_{-L/2}^{L/2} \mathrm{d}x~ \sqrt{\frac{r^6}{l^6} f(r) \left(1+ \frac{l^4 r'^2}{r^4 f(r)}\right)}= \int_{-L/2}^{L/2} \mathrm{d}x~\sqrt{F^2+ G^2 r'^2},\nonumber   
\end{equation}
which is universal to all the backgrounds and QFTs studied here. The treatment of this variational problem is analogous to the one we presented for the Wilson loop. 

{
We refer the reader to \cite{Kol:2014nqa} for a careful derivation of the integral expressions for the length of the strip and the EE, both in terms of the turning point $r_0$. 
To be self-contained, we present some details of the extremisation procedure. From the above expression, we have
\begin{eqnarray}
 & & S_{EE}= \frac{{\cal N}}{4 G_N}\int_{-L/2}^{L/2} dx \sqrt{F^2+ G^2 r'^2},~~~F^2=\frac{r^6}{l^6}f(r),~~G^2= \frac{r^2}{l^2}\label{ee1}
\end{eqnarray}
The conserved Hamiltonian for this system is
\begin{equation}
    H= \frac{F^2}{\sqrt{F^2+G^2 r'^2}},
    \end{equation}
    equating the Hamiltonian to the (conserved) value $H= F(r_0)$, where $r_0$ is the turning point of the surface, we find the expressions,
\begin{equation}
 dx=\frac{F(r_0) G(r)}{F(r)}\frac{dr}{\sqrt{F^2(r)- F^2(r_0)}},~~~L(r_0)= 2 F(r_0)\int_{r_0}^\infty dr \frac{G(r)}{F(r)}\frac{dr}{\sqrt{F^2(r)- F^2(r_0)}}.  
\end{equation}    
A regularised expression for the Entanglement Entropy is obtained replacing the expression for $r'$ in eq.(\ref{ee1}),
\begin{equation}
 S_{EE}=\frac{{\cal N}}{4 G_N} \Big[ \int_{r_0}^\infty dr \frac{G(r) F(r)}{\sqrt{F^2(r)- F^2(r_0)}}  - \int_{r_*}^\infty dr G(r)\Big].\label{ee2}
\end{equation}
}
Using these integral expressions we numerically evaluate $L_{\text{EE}}(r_0)$, $S_{\text{EE}}(r_0)$ and parametrically plot $S_{\text{EE}}(L_{\text{EE}})$.
In Figure \ref{EEfigure}, we display the results for $L_{\text{EE}}(r_0)$ together with the approximate expression, in the same line as eq.(\ref{Lapp-Wilson}). The double valued character of $L_{\text{EE}}(r_0)$ indicates that there is a phase transition in the EE, as proposed in \cite{Klebanov:2007ws}. Figure \ref{EEfigure} shows also the EE in terms of $L_{\text{EE}}$, where the phase transition is more obvious. As with the Wilson loop, a more general embedding might reveal different aspects of the dynamics.

\subsubsection{Flow central charge}
The goal of this section is to write a quantity that measures the degrees of freedom of the QFTs in question. This must be monotonous, indicate the IR gapped character of the QFTs (i.e. vanish towards the IR) and detect the presence of the conformal fixed point in the UV (i.e. asymptote to a constant value characteristic of the CFT). This is not an easy task because the QFTs are flowing across dimensions--- we have $(2+1)$ QFTs in the IR and $(3+1)$ CFTs in the far UV. In other words, these are non-isotropic QFTs. Fortunately, holography can help us. In \cite{Bea:2015fja}, see also \cite{Merrikin:2022yho}, a generalisation of the holographic central charge is introduced called $c_{flow}$. This quantity is capable of detecting fixed points along flows across dimensions. We briefly summarise the idea here.

Consider a background conjectured as dual to a $(d+1)$-dimensional QFT. The metric and dilaton are of the form,
\begin{eqnarray}
& & \mathrm{d}s^2=-\alpha_0 \mathrm{d}t^2 +\alpha_1 \mathrm{d}x_1^2+ \alpha_2 \mathrm{d}x_2^2+....+ \alpha_d \mathrm{d}x_d^2 +\left(\alpha_1 .\alpha_2....\alpha_d\right)^{\frac{1}{d}}\beta(r) \mathrm{d}r^2+ g_{ij}(\mathrm{d}\theta^i-A^i)(\mathrm{d}\theta^j-A^j),\nonumber\\
& &\Phi=\Phi(r,\theta^i).\label{flow-back}
\end{eqnarray}
We define,
\begin{eqnarray}
& &G_{ij}\mathrm{d}\xi^i \mathrm{d}\xi^j= \alpha_1 \mathrm{d}x_1^2+ \alpha_2 \mathrm{d}x_2^2+....+ \alpha_d \mathrm{d}x_d^2 + g_{ij}(\mathrm{d}\theta^i-A^i)(\mathrm{d}\theta^j-A^j),\nonumber\\
& & H= \biggl[ \int \mathrm{d}\theta^i \sqrt{e^{-4\Phi} \det[ G_{ij}]}\biggr]^2,~~~~~~~~~~~
 c_{flow}= \frac{d^d}{G_N} \beta^{\frac{d}{2}} \frac{H^{\frac{2d+1}{2}}}{\left( H' \right)^d}.\label{cflow}
\end{eqnarray}
In the case of duals to CFTs (i.e. isotropic systems with equal $\alpha_i$) the result of c$_{flow}$ coincides with the free energy of the CFT. Various checks for SCFTs in diverse dimensions can be found in 
\cite{Nunez:2023loo}. For the cases that occupy us here, we have field theories in $d=3$. The values of the $\alpha_i$, $\beta(r)$, $H(r)$  are,
\begin{eqnarray}
& & \underline{\text{case of}~\bf S^5}: \alpha_1=\alpha_2= \frac{r^2}{l^2},~\alpha_3= \frac{r^2}{l^2}f(r),~~\beta= \frac{l^4}{r^4 f(r)^{\frac{4}{3}}},~~~H= \frac{{\cal N}_{S^5}^2}{L_\phi^2 L_{x_2}^2} \frac{r^6}{l^6}f(r), \label{cantidades} \\
& & \underline{\text{case of}~\bf Y^{p,q}}: \alpha_1=\alpha_2= \frac{r^2}{l^2},~\alpha_3= \frac{r^2}{l^2}f(r),~~\beta= \frac{l^4}{r^4 f(r)^{\frac{4}{3}}},~~~H= \frac{{\cal N}_{Y^{p,q}}^2}{L_\phi^2 L_{x_2}^2} \frac{r^6}{l^6}f(r).\nonumber\\
& & \underline{\text{case of}~\bf GM}: \alpha_1=\alpha_2= \sqrt{f_1^3 f_5} \frac{4 r^2}{l^2},~\alpha_3= \alpha_1 f(r),~~\beta= \frac{l^4}{r^4 f(r)^{\frac{4}{3}}},~~~H= \frac{{64\cal N}_{GM}^2}{L_\phi^2 L_{x_2}^2} 
\frac{ r^6}{l^6}f(r).\nonumber\\
& & \underline{\text{case of}~\bf BPT}: \alpha_1=\alpha_2= 18 \pi \sqrt{-\frac{\alpha}{\ddot{\alpha}} } \frac{r^2}{l^2},~\alpha_3= \alpha_1 f(r),~~\beta= \frac{l^4}{r^4 f(r)^{\frac{4}{3}}},~~~H= \frac{{\cal N}_{BPT}^2}{L_\phi^2 L_{x_2}^2} \frac{r^6}{l^6}f(r).\nonumber
\end{eqnarray}
We have used the quantities ${\cal N}_i$ defined in eq.(\ref{EEdiego}), for $i \in \{S^5, \mathrm{Y}^{p,q}, GM, BPT\}$. Using eq.(\ref{cflow}) we find,
\begin{eqnarray}
 c_{flow}= \frac{s_i l^3 {\cal N}_i}{8 L_\phi L_{x_2} G_N} \frac{f(r)^{\frac{3}{2}}}{\left(f(r)+\frac{r}{6}f'(r) \right)^3}.\label{cflowresult}   
\end{eqnarray}
Where $s_i=8$ for the GM-case, and is $s_i=1$ otherwise.

Similar comments as those made for the EE apply here. 
It is clear from these results that c$_{flow}$ receives a contribution from the UV, represented by ${\cal N}_i$, which is related to the number of degrees of freedom of each CFT$_4$. The contribution from the flow is ruled by the $r$-dependent factor. Note that for $\mu=q=0$ the function $f(r)=1$ and c$_{flow}= c_{CFT_4}$.
Also, for $r\to r_{*}$ we have c$_{flow}\to 0$, indicating the gapped character of the IR QFT$_3$.

In summary, we have found a monotonic quantity that interpolates between the number of degrees of freedom for a CFT$_4$ in the UV (large $r$) and a gapped QFT$_3$ in the IR (small $r$).

\subsubsection{Other observables}
Let us close our investigation of holographic observables by briefly studying some probe branes that are more sensitive to the particularities of the deformed QFT. We are very succinct as this is analysed more detail in the companion paper \cite{Chatzis:2024kdu}.

Let us start by considering a probe D7 brane, that in the geometries of eqs.(\ref{metric-AdS5xS5}) and (\ref{upliftYpq}) extend over ${\cal M}_8=[t,x_1,x_2,\theta,\varphi,\varphi_1,\varphi_2,\varphi_3]$ at constant values  $r=r_{*},\phi=\phi_0$. 
We also allow a gauge field $a_1$ on the brane worldvolume, with field strength $f_2=\mathrm{d}a_1$.

Our main interest is on the WZ part of the action for this D7 probe,
\begin{eqnarray}
S_{WZ}=-T_{D7}\int C_8 - C_5\wedge f_2 +\frac{1}{2}C_4\wedge f_2\wedge f_2.    
\end{eqnarray}
In our backgrounds $C_8=C_5=0$. Integrating the remaining term by parts,
\begin{equation}
 S_{WZ}= \frac{T_{D7}}{2}\int \mathrm{d}^{2+1}x \, a_1\wedge f_2\int_{S^5}F_5=   \frac{T_{D7} 16 \pi^4 g_s \alpha'^2}{2} N_{D3}\int \mathrm{d}^{2+1}x~  a_1\wedge f_2  ,
\end{equation}
where we have used eq.(\ref{d3quantised}). Note that this action suggests that a Chern-Simons theory with level $N_{D3}$ is present in our holographic system. The calculation is exactly the same in the $\mathrm{Y}^{p,q}$ case.

The situation is analogous, although with subtle differences, for the case of the background in eqs.(\ref{metric-dil-BPT})-(\ref{RR-NS-BPT}). Indeed, we probe the background in eq.(\ref{metric-dil-BPT}) with a  D6 brane that extends on ${\cal M}_7=[t,x_1,x_2, z, v_1,v_2,\psi]$ at constant values  $\theta=\frac{\pi}{2},\phi=\phi_0, r=r_{*}$. Using that the magnetic parts of $C_7,C_5$ vanish in the background, we find after an integration by parts that the WZ term is given by
\begin{equation}
 S_{WZ}=\frac{T_{D6}}{2}  \int_{\psi,z, \Sigma}\mathrm{d} C_3 ~ \int \mathrm{d}^{2+1} x ~a_1\wedge f_2.\label{CSmassiveiia}
\end{equation}
We have used that on this submanifold
\begin{equation}
  \xi=0,~~ C_1=B_2={\cal F}_2=0,~~D\psi=\mathrm{d}\psi,~~~F_4= \mathrm{d}C_3=\frac{\ddot{\alpha}}{162\pi} \mathrm{d}z\wedge \text{Vol}_{\Sigma}\wedge \mathrm{d}\psi.\nonumber  
\end{equation}
Using now eq.(2.12) in \cite{Filippas:2019puw} we find,
\begin{equation}
 S_{WZ}= \frac{T_{D6}}{2} N_{D6}\text{Vol}_{\Sigma} \int \mathrm{d}^{2+1} x ~a_1\wedge f_2,
\end{equation}
where $N_{D6}$ is the total number of D6 branes in the six dimensional quiver, which under compactification on $\Sigma_2$ yields the 4d non-Lagrangian ${\cal N}=1$ SCFT (in the UV) that we deformed to a 3d QFT in the IR. 

One can carry out this calculation for the background in eq.(\ref{eq:N=2}) using a probe D6 extended on $[t,x_1,x_2,\eta, \chi, \theta,\varphi]$. Calculating the integer number appearing in the flux of $F_4$ is more subtle in this case. We discuss this analysis in more detail in \cite{Chatzis:2024kdu}. In summary, we observe that level $N$ Chern-Simons actions appear in our systems.

Based on the arguments of \cite{Aharony:1998qu}, one expects the QFT$_3$ to have various vacua separated by domain walls. These domain walls should be understood as D1 branes stretched along $(t,x_1)$ and at the point $r=r_{*}$. The D1s would have constant tension in all the backgrounds we have discussed. It should be interesting to understand the origin of these discrete vacua holographically. One should be able to apply the logic of \cite{Aharony:1998qu}, this applies in our Type IIB backgrounds straightforwardly. The application to set-ups based on Type IIA is less clear. We leave this for future study.

\section*{Acknowledgments} For discussions, comments on the manuscript and for sharing their ideas with us, we wish to thank: Davide Cassani, Alejandra Castro, S. Prem Kumar, Leonardo Santilli, Ricardo Stuardo, Oscar Varela, Luigi Tizzano, David Turton. We are supported by the grants ST/Y509644-1, ST/X000648/1 and ST/T000813/1. The work of AF has been supported by the STFC Consolidated  Grant ST/V507143/1 and by the EPSRC Standard Research Studentship (DTP)  EP/T517987/1.


{\bf Open Access Statement}---For the purpose of open access, the authors have applied a Creative Commons 
Attribution (CC BY) licence to any Author Accepted Manuscript version arising.

\bibliographystyle{JHEP}
\bibliography{Ref.bib}

\end{document}